\documentclass[preprint,showpacs,12pt,epsfig,floatfix,aps]{revtex4}
 
\usepackage{graphicx}
\usepackage{latexsym}
\usepackage{amsmath}
\usepackage{epstopdf}
\input psfig.sty

\newcommand{\bq}{\begin{equation}}
\newcommand{\eq}{\end{equation}}
\newcommand{\bqn}{\begin{eqnarray}}
\newcommand{\eqn}{\end{eqnarray}}
\newcommand{\nb}{\nonumber}
\newcommand{\lb}{\label}
\newcommand{\rr}{\bf r}

\begin{document}
\title{Radiating Gravastars}
\author{R. Chan $^{1}$}
\email{chan@on.br}
\author{M.F.A. da Silva $^{2}$}
\email{mfasnic@gmail.com}
\author{Jaime F. Villas da Rocha $^3$}
\email{jfvroch@pq.cnpq.br}
\author{Anzhong Wang $^{4}$}
\email{anzhong_wang@baylor.edu}
\affiliation{\small $^{1}$ Coordena\c{c}\~ao de Astronomia e Astrof\'{\i}sica, 
Observat\'orio Nacional, Rua General Jos\'e Cristino, 77, S\~ao Crist\'ov\~ao  
20921-400, Rio de Janeiro, RJ, Brazil\\
$^{2}$ Departamento de F\'{\i}sica Te\'orica, Instituto de F\'{\i}sica, 
Universidade do Estado do Rio de Janeiro, Rua S\~ao Francisco Xavier 524, 
Maracan\~a 20550-900, Rio de Janeiro - RJ, Brasil\\
$^{3}$ Universidade Federal do Estado do Rio de Janeiro,
Instituto de Bioci\^encias,
Departamento de Ci\^encias Naturais, Av. Pasteur 458, Urca,
CEP 22290-240, Rio de Janeiro, RJ, Brazil \\
$^{4}$ GCAP-CASPER, Department of Physics,
Baylor University, Waco, TX 76798, USA}

\date{\today}

\begin{abstract}
Considering a Vaidya exterior spacetime, we study dynamical models of
prototype gravastars, made of an infinitely thin spherical shell of a
perfect fluid with the equation of state $p = \sigma$, enclosing an 
interior de Sitter spacetime. 
We show explicitly that the final output can be a black
hole, an unstable gravastar, a stable gravastar or a "bounded excursion" 
gravastar, depending on how the mass of the shell evolves in time, the 
cosmological constant and the initial position of the dynamical shell.
This work presents, for the first time in the
literature, a gravastar that emits radiation.
\end{abstract}


\maketitle

\section{Introduction}

Gravastar was proposed as an alternative to black holes.
The initial model  of Mazur and Mottola (MM) \cite{MM01},
consists of five layers: an internal core
$0 < r < r_1$, described by the de Sitter universe, an intermediate thin layer of stiff fluid
$r_1 < r < r_2$, an external region $r > r_2$, described by the Schwarzschild solution, and two 
infinitely thin shells, appearing, respectively, on the hypersurfaces $r = r_1$ and
$r = r_2$. The intermediate layer is constructed in such way that $r_1$ is inner than the de Sitter horizon, 
while $r_2$ is outer than the Schwarzschild horizon, eliminating the existence of any  horizon. Configurations with 
a de Sitter interior have long history which we can find, for example, in the work of Dymnikova and Galaktionov \cite{irina}.  
After this work, Visser and Wiltshire \cite{VW04} (VW) pointed out that there are 
two different types of stable gravastars which are stable gravastars and 
"bounded excursion" gravastars. In the spherically symmetric case, the motion 
of the surface of the gravastar can be written in the form \cite{VW04},
\bq
\lb{1.4}
\frac{1}{2}\dot{R}^{2} + V(R) = 0,
\eq
where $R$ denotes the radius of the star, and $\dot{R} \equiv dR/d\tau$, with 
$\tau$ being the proper time of the surface. Depending on the properties of 
the potential $V(R)$, the two kinds of gravastars are defined as follows. 

{\bf Stable gravastars}: In this case,  there must exist a radius $a_{0}$ such that
\bq
\lb{1.5}
V\left(R_{0}\right) = 0, \;\;\; V'\left(R_{0}\right) = 0, \;\;\;
V"\left(R_{0}\right) > 0,
\eq
where a prime denotes the ordinary differentiation with respect to the indicated argument.
If and only if there exists such a radius $R_{0}$ for which the above conditions are satisfied,
the model is said to be stable. 
Among other things, VW found that there are many equations of state for which the gravastar
configurations are stable, while others are not \cite{VW04}. Carter studied  the same
problem and found new equations of state for which the gravastars are stable \cite{Carter05}, 
while De Benedictis {\em et al} \cite{DeB06} and Chirenti and Rezzolla \cite{CR07} 
investigated the stability of the original model
of  Mazur and  Mottola against axial-perturbations, and found that gravastars are stable to
these perturbations,  too. Chirenti and Rezzolla also showed that their quasi-normal modes 
differ from those of black holes with the same mass, and thus can be used to discern a gravastar 
from a black hole. 

{\bf "Bounded excursion" gravastars}: As VW noticed, there is a less stringent notion of 
stability, the so-called "bounded excursion" models, in which there exist two radii $a_{1}$ 
and $a_{2}$ such that
\bq
\lb{1.6}
V\left(R_{1}\right) = 0, \;\;\; V'\left(R_{1}\right) \le 0, \;\;\;
V\left(R_{2}\right) = 0, \;\;\; V'\left(R_{2}\right) \ge 0,
\eq
with $V(R) < 0$ for $R \in \left(R_{1}, R_{2}\right)$, where $R_{2} > R_{1}$. 

Lately, we studied both types of gravastars \cite{JCAP}-\cite{JCAP5}, and found that, 
such configurations can indeed be constructed, although   the region for the formation 
of them in the phase space  is very small in comparison to that of black holes.

Based on the discussions about the gravastar picture some authors have
proposed alternative models \cite{Chan}-\cite{Lobo}. In addition, since in 
the study
of the evolution of gravastar there is a possibility of black hole formation,
we can find some works considering the hypothesis of dark energy black hole
\cite{Debnath}\cite{Chan}\cite{Cai}\cite{Bronnikov1}\cite{Bronnikov2}.

In the last four years we have adopted a different approach (from VW \cite{VW04}), 
which means that we started from an equation of state and found
the potential of the shell \cite{JCAP}\cite{JCAP1}. Generalizing the exterior spacetime in
order to include a cosmological constant, we study the gravastar model in a
de Sitter-Schwarzschild spacetime, which allowed to  investigate the role of
cosmological constant in its evolution \cite{JCAP3}. Following this direction, we
also considered a de Sitter-Reissner-Nordstr\"om exterior spacetime \cite{JCAP4}. On
the other hand, we studied the effects of changing the interior of the
gravastar, filling it with  anisotropic fluids, which can be characterized
by different kinds of dark energy \cite{JCAP2}\cite{JCAP5}.

Here we are interested in the study of a gravastar model whose
interior consists of a de Sitter spacetime and an exterior
radiative Vaidya's spacetime.
The paper is  organized as follows: In Sec. II we present the metrics of the 
interior and exterior spacetimes, and write down the motion of the thin shell
in the form of  equation (\ref{1.4}).  In Sec. III we study the model by using  the small radiating 
source approximation.  In Sec. IV  we discuss the formation of
black holes and  gravastars, when the mass of the thin shell increases, while in Sec. V we study the case
where the mass of the thin shell decreases. 
Finally, in Sec. VI we present our main conclusions.

\section{ Dynamical Three-layer Prototype Gravastars}

The interior spacetime is described by the de Sitter's metric given by
\bq
ds^2_{i}=-f dt^2 + f^{-1} dr^2 + r^2 d\Omega^2,
\lb{ds2-}
\eq
where $f=1- (r/L)^2$, $L=\sqrt{3/\Lambda}$ and
$d\Omega^2 = d\theta^2 + \sin^2(\theta)d\phi^2$.

The exterior spacetime is given by the Vaidya's metric
\bq
ds^2_{e}= - F dv^2 - 2 d{\rr} dv + {\rr}^2 d\Omega^2,
\lb{ds2+}
\eq
where $F=1 - \frac{2m(v)}{\rr}$.
The metric of the hypersurface  on the shell is given by
\bq
ds^2_{\Sigma}= -d\tau^2 + R^2(\tau) d\Omega^2,
\lb{ds2Sigma}
\eq
where $\tau$ is the proper time.

Since $ds^2_{i} = ds^2_{e} = ds^2_{\Sigma}$, we find that  $r_{\Sigma}={\rr}_{\Sigma}=R$,
and  
\bqn
\lb{dott2}
f\dot t^2 - f^{-1} \dot R^2 &=&  1,\\
\lb{dotv2}
\left[ F+\frac{2\dot R}{\dot v} \right] \dot v^2 &=& 1,
\eqn
where the dot denotes the ordinary differentiation with respect to the proper time.
On the other hand,  the interior and exterior normal vectors to the thin shell are given by
\bqn
\lb{nalpha-}
n^{i}_{\alpha} &=& (-\dot R, \dot t, 0 , 0 ),\nb\\
n^{e}_{\alpha} &=& (-\dot R, \dot v, 0 , 0 ).
\eqn
Then, the interior and exterior extrinsic curvatures are given by
\bqn
K^{i}_{\tau\tau}&=&-[(3 L^4 \dot R^2-L^4 \dot t^2+2 L^2 R^2 \dot t^2-
R^4 \dot t^2) R \dot t-(L+R) (L-R) (\dot R \ddot t-\ddot R \dot t) L^4] \times \nb \\
& &(L+R)^{-1} (L-R)^{-1} L^{-4}
\lb{Ktautau-}
\eqn
\bq
K^{i}_{\theta\theta}=\dot t (L+R) (L-R) L^{-2} R
\lb{Kthetatheta-}
\eq
\bq
K^{i}_{\phi\phi}=K^{i}_{\theta\theta}\sin^2(\theta),
\lb{Kphiphi-}
\eq
\bq
K^{e}_{\tau\tau} = \dot v^2(2 m^2 \dot v-3 m R \dot R-m R \dot v+
\dot m R^2 \dot v) R^{-3},
\lb{Ktautau+}
\eq
\bq
K^{e}_{\theta\theta} = -\dot v (2 m-R)+R \dot R,\\
\lb{Kthetatheta+}
\eq
\bq
K^{e}_{\phi\phi} = K^{e}_{\theta\theta}\sin^2(\theta).
\lb{Kphiphi+}
\eq
Since \cite{Lake}
\bq
[K_{\theta\theta}]= K^{e}_{\theta\theta}-K^{i}_{\theta\theta} = - M,
\lb{M}
\eq
where $M$ is the mass of the shell, we find that
\bq
M=\dot v (2 m-R)+\dot t(1-2a R^2) R.
\lb{M1}
\eq
Then, substituting equations (\ref{dott2}) and (\ref{dotv2}) into (\ref{M1}) 
we get
\bq
M=-R \dot R + \left(2m-R \right) \dot v + 
R\left[ \dot R^2  + 1 - \left( \frac{R}{L} \right)^2 \right]^{1/2}.
\lb{M2}
\eq
In order to keep the ideas of MM as much as possible, we consider the thin 
shell as consisting
of a fluid with the equation of state, $p=\sigma$, where $\sigma$ and $p$ denote, 
respectively, the surface energy density and pressure of the shell. 
Then, the equation of motion of the shell is given by \cite{Lake}
\bq
\dot M + 8\pi R \dot R p = 4 \pi R^2 [T_{\alpha\beta}u^{\alpha}n^{\beta}]=
4\pi R^2 \left(T^e_{\alpha\beta}u_e^{\alpha}n_e^{\beta}-T^i_{\alpha\beta}u_i^{\alpha}n_i^{\beta} \right),
\lb{dotM}
\eq
where $u^{\alpha}$ is the four-velocity.  

The exterior energy-momentum tensor is given by
\bq
T^e_{\alpha\beta}=\epsilon l_\alpha l_\beta
\eq
where 
\bq
l_\alpha= \delta_\alpha^v,\;\;\; l_\alpha l^\alpha=0, 
\eq
\bq
\epsilon=-\frac{\dot v^2}{4\pi R^2}\frac{dm}{dv}.
\eq
 
Since
\bq
u^e_\alpha=(\pm(\dot R+F\dot v), \pm\dot v, 0,0),
\eq
(in this paper we shall choose the plus signal), we find 
\bq
T^e_{\alpha\beta}u_e^{\alpha} n_e^{\beta}= - \frac{\dot m \dot v^3}{4\pi R^2}. 
\eq
Since the interior spacetime is de Sitter, we get
\bq
\dot M + 8\pi R \dot R \sigma = -\dot m \dot v^3.
\lb{dotM1}
\eq

In order to solve this equation, let us assume that
\bq
\dot m \dot v^3 = k_0 = k_1 \dot M,
\lb{k0}
\eq
where $k_0$ and $k_1$ are constant.  Note that $k_0<0$ since $\epsilon>0$.
Thus, recalling that $\sigma = M/(4\pi R^2)$, 
we  find that the solution of equation (\ref{dotM1}) is given by, 
\bq
M=k R^{-\frac{2}{k_1+1}},
\lb{Mk}
\eq
where $k$ is a positive integration constant. Note also that $k_1=-1$ is not allowed.
If $k_1 \ll 1$ we get the approximation of small emission of radiation.
If $k_1 < 0$ then from equation (\ref{k0}) we can see that $\dot M > 0$, meaning that
the mass of the shell is increasing.  In this case, the interior de Sitter energy  transfers 
to   the thin shell, feeding the outgoing radiation.
If $k_1 > 0$ then from equation (\ref{k0}) we can see that $\dot M < 0$, meaning that
the mass of the shell is decreasing. In this case, the thin shell looses mass in order
to maintain the outgoing radiation. 
Thus, for three particular values of $k_1$, which cover all the
possibilities described above, we have

\bq
M=\left\{
\begin{array}{ll}
k R^{-2}, & \mbox{if taking the limit $k_1 \rightarrow 0$, } \\
k R^{+2}, & \mbox{if $k_1=-2$,} \\
k R^{-1}, & \mbox{if $k_1=+1$,}
\end{array}
\right.
\eq
which will be studied in more details in the next three sections.

Substituting equation (\ref{Mk}) into equation (\ref{M2}),
we find that  the potential in the form of  equation (\ref{1.4}) can be written as
\bqn
& &V(R,m,L,k)= -\frac{1}{8L^4 R^2 k^2 R^{-\frac{4}{k_1+1}}} \left( -4 R^5 m L^2-4 L^4 R^2 k^2 R^{-\frac{4}{k_1+1}}+ \right. \nb \\
& & \left. 2 R^4 k^2 R^{-\frac{4}{k_1+1}} L^2+4 R^2 m^2 L^4+4 R m L^4 k^2 R^{-\frac{4}{k_1+1}}+R^8+k^4 R^{-\frac{8}{k_1+1}} L^4 \right). 
\lb{VRmL}
\eqn

Rescaling $m, \; L$ and $R$ as,
\bqn
m &\rightarrow& mk^{\frac{k_1+1}{k_1+3}},\nb\\
L &\rightarrow& Lk^{\frac{k_1+1}{k_1+3}},\nb\\
R &\rightarrow& Rk^{\frac{k_1+1}{k_1+3}},
\eqn
we find that equation (\ref{VRmL}) can be written as 
\bqn
& &V(R,m,L)= -\frac{1}{8L^4 R^2} \times \nb \\
& & \left( -4 R^{\frac{9+5 k_1}{k_1+1}} m L^2-4 L^4 R^2+2 R^4 L^2+4 R^{2 \frac{3+k_1}{k_1+1}} m^2 L^4+4 R m L^4+R^{4 \frac{3+2 k_1}{k_1+1}}+R^{-\frac{4}{k_1+1}} L^4\right). \nb \\
\lb{VR}
\eqn
Clearly, for any given constants $m$ and $L$, equation (\ref{VR}) uniquely 
determines the collapse of the prototype  gravastar. Depending on the initial value $R_{0}$,  
the collapse can form either a black hole,  a gravastar or 
a de Sitter spacetime. In the last case, the thin shell
first collapses to a finite non-zero minimal radius and then expands to infinity.  To  guarantee
that initially the spacetime does not have any kind of horizons,  cosmological or event,
we must restrict $R_{0}$ to the range,
\bq
\lb{2.2b}
2m < R_{0} < L,
\eq
where $R_0$ is the initial collapse radius.

\section{Gravastars/Black Holes with  Small Emission of Radiation}

Here we will consider the limit $k_1 \rightarrow 0$. Then, we find that 
\bq
V(R,m,L) = -\frac{1}{8L^4 R^6} \left(-4 L^4 R^6+2 R^8 L^2+R^{16}-4 R^{13} m L^2+4 R^{10} m^2 L^4+4 R^5 m L^4+L^4\right),
\lb{VR0}
\eq
fro which we find that   $\lim_{R\rightarrow 0} V(R,m,L) = \lim_{R\rightarrow \infty} V(R,m,L) =-\infty$.

The first derivative of the potential, equation (\ref{VR0}), is given by
\bq
\frac{dV}{dR}= -\frac{1}{4L^4 R^7} \left(2 R^8 L^2+5 R^{16}-14 R^{13} m L^2+8 R^{10} m^2 L^4-2 R^5 m L^4-3 L^4\right).
\eq

Thus the solutions for $\frac{dV}{dR}=0$ are
\bq
m_1 = \frac{1}{8R^5 L^2}\left(7 R^8+L^2+\sqrt{9 R^{16}-2 R^8 L^2+25 L^4}\right),
\lb{m1}
\eq
and
\bq
m_2 = \frac{1}{8R^5 L^2}\left(7 R^8+L^2-\sqrt{9 R^{16}-2 R^8 L^2+25 L^4}\right).
\lb{m2}
\eq

\begin{figure}
\vspace{.2in}
\centerline{\psfig{figure=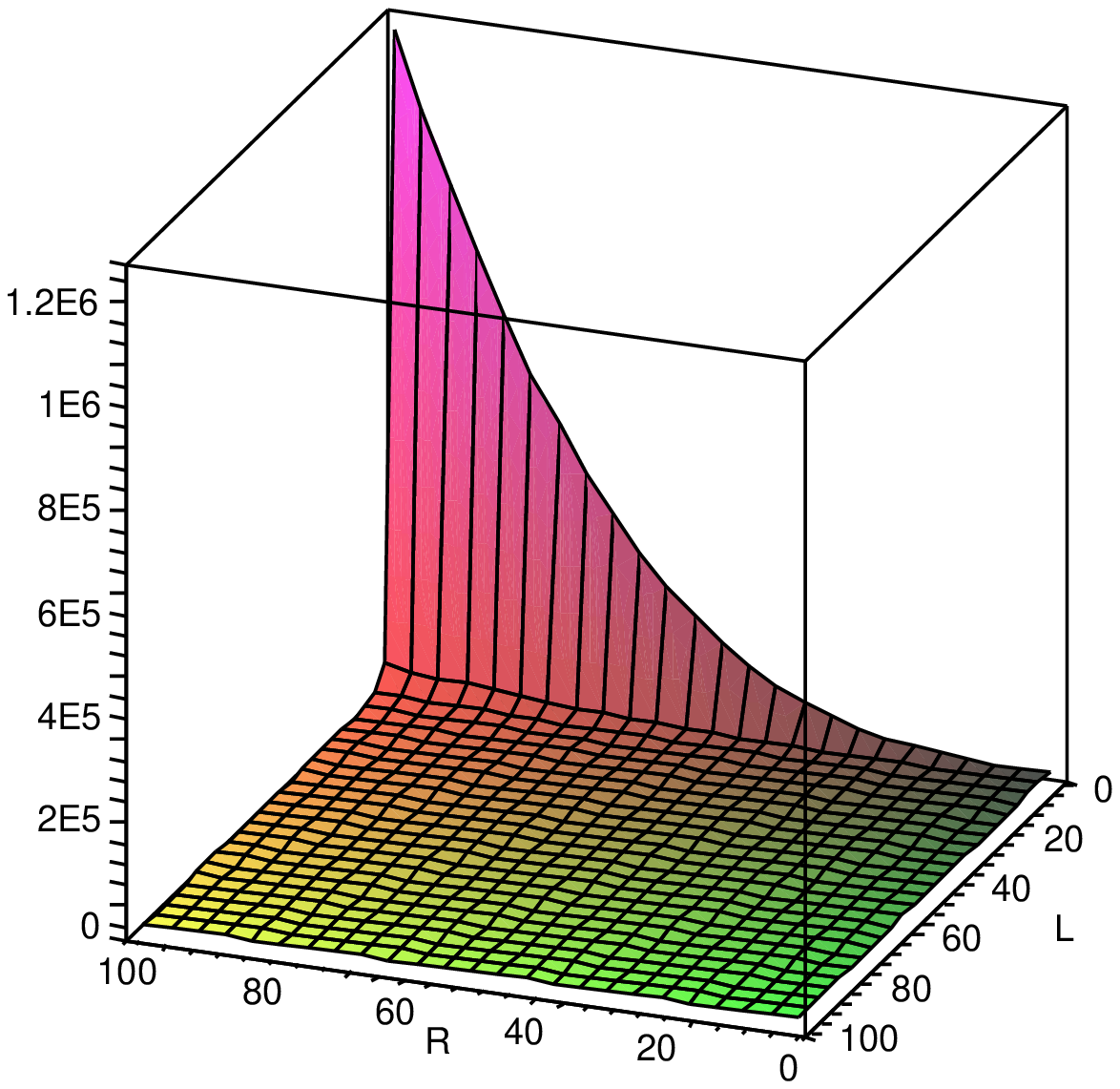,width=3.0truein,height=3.0truein}\hskip
.1in \psfig{figure=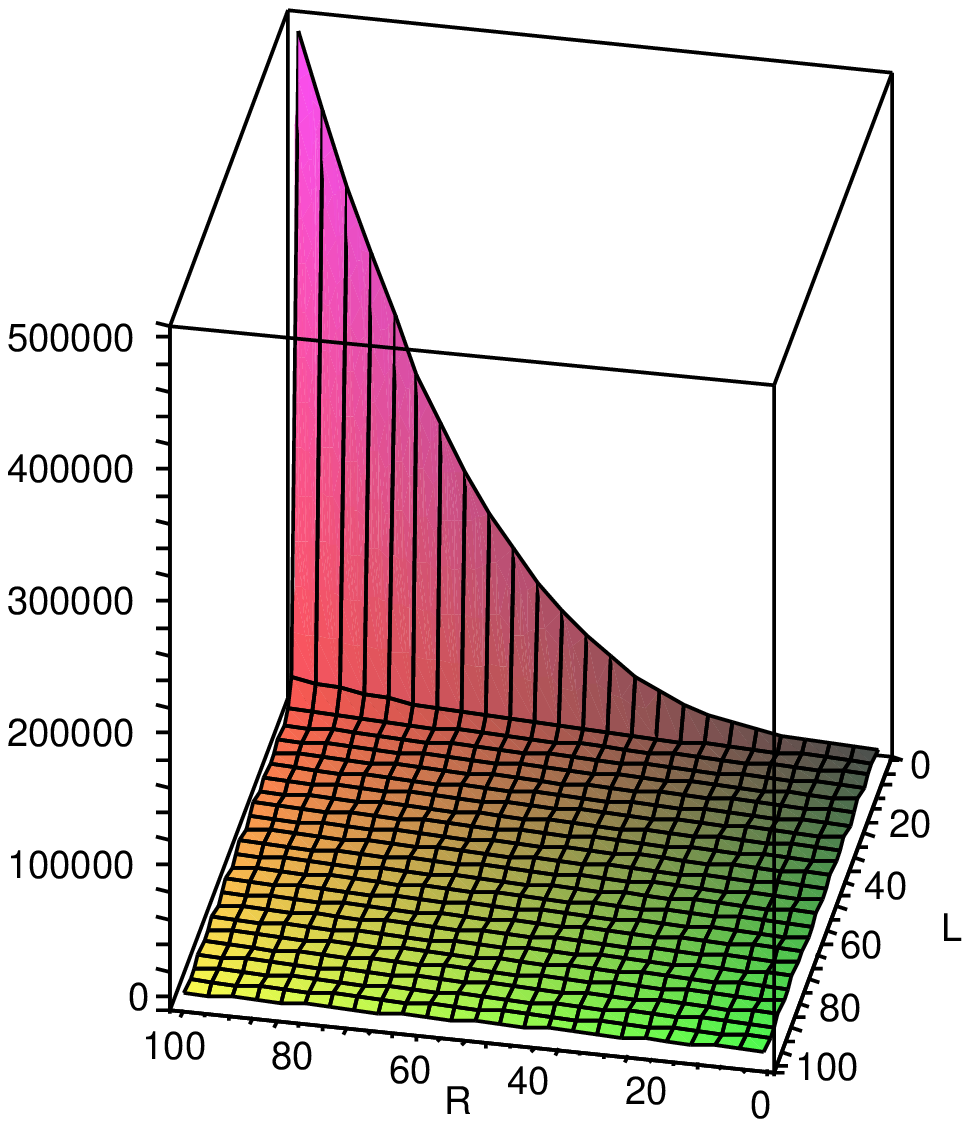,width=3.0truein,height=3.0truein}
\hskip .5in} \caption{Case $k_1 \simeq 0$. The masses $m_1$ (left) and $m_2$ (right) where the first derivative of
the potential $V(R)$ is zero.  We note that both masses are positive.}
\label{fig1casek10}
\end{figure}

Note that  $m_2$ and $m_1$ are always positive, as can be seen from  Fig. \ref{fig1casek10}.

The second derivative of the potential is given by
\bq
\frac{d^2V}{dR^2}(R,m,L)= -\frac{1}{4R^8 L^4} \left(2 R^8 L^2+45 R^{16}-84 R^{13} m L^2+24 R^{10} m^2 L^4+4 R^5 m L^4+21 L^4 \right).
\lb{VR2}
\eq

\begin{figure}
\vspace{.2in}
\centerline{\psfig{figure=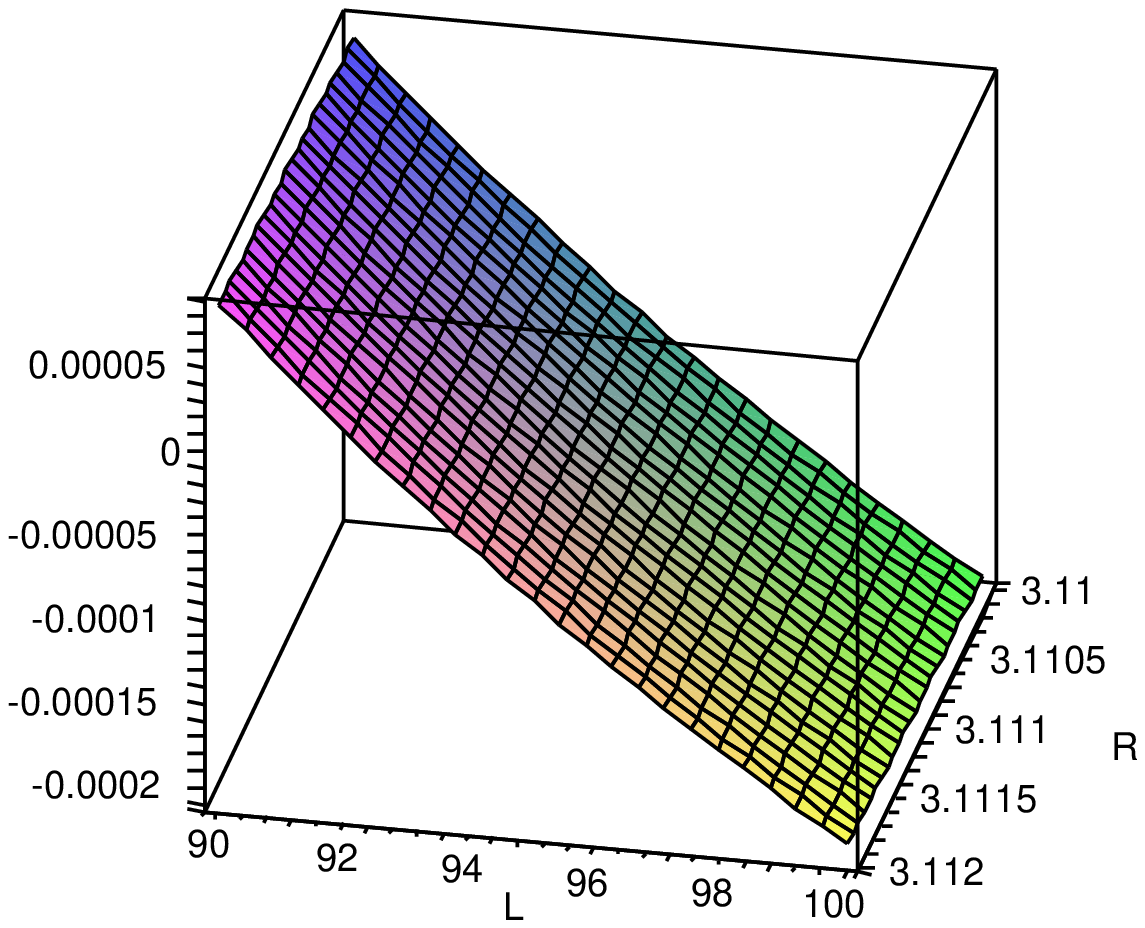,width=3.0truein,height=3.0truein}
\hskip .05in \psfig{figure=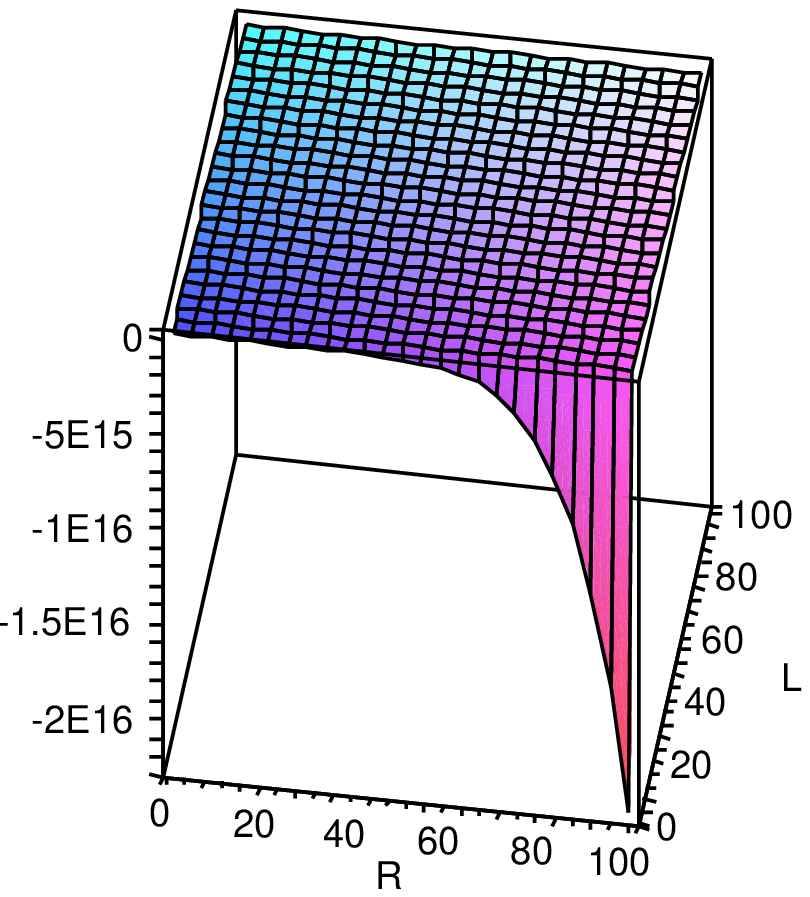,width=3.0truein,height=3.0truein}
\hskip .5in} \caption{Case $k_1 \simeq 0$. The second derivative of the potential 
$\frac{d^2V}{dR^2}(R,m,L)$ calculated at $m=m_1$ (left) and $m=m_2$ (right).
We  note that $\frac{d^2V}{dR^2}(R,m=m_1,L)$ can be positive or negative
(the frontier between the two regions is given by equation (\ref{Lf1}))
and that $\frac{d^2V}{dR^2}(R,m=m_2,L)$ is always negative.}
\label{fig2casek10}
\end{figure}

\begin{figure}
\vspace{.2in}
\centerline{\psfig{figure=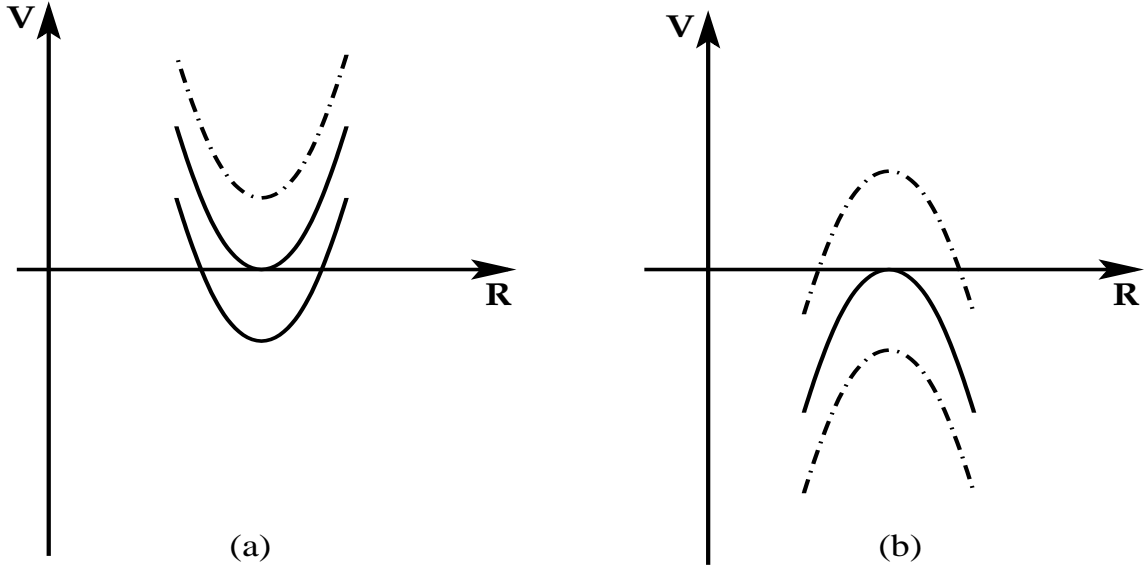,width=6truein,height=3.0truein}
\hskip .5in} \caption{The possible type of potentials.}
\label{potenciais}
\end{figure}

Substituting equation (\ref{m1}) into equation (\ref{VR2}) we have
\bqn
& &\frac{d^2V}{dR^2}(R,m=m_1,L)=-\frac{1}{16L^4R^8} \times \nb \\
& &\left(-2 R^8 L^2-27 R^{16}-21 R^8 \sqrt{9 R^{16}-
2 R^8 L^2+25 L^4}+125 L^4+5 L^2 \sqrt{9 R^{16}-2 R^8 L^2+25 L^4}\right). \nb \\
\eqn

Solving $\frac{d^2V}{dR^2}(R,m=m_1,L)=0$ we get
\bq
L_f^2 \approx  0.9731857906 R^8.
\lb{Lf1}
\eq

Substituting equation (\ref{m2}) into equation (\ref{VR2}) we have
\bqn
& &\frac{d^2V}{dR^2}(R,m=m_2,L)=-\frac{1}{16L^4R^8} \times \nb \\
& &\left(-2 R^8 L^2-27 R^{16}+21 R^8 \sqrt{9 R^{16}-
2 R^8 L^2+25 L^4}+125 L^4-5 L^2 \sqrt{9 R^{16}-2 R^8 L^2+25 L^4}\right). \nb \\
\eqn

Thus, we can see from Fig. \ref{fig2casek10} that the second derivative 
of the potential is always positive at $m=m_1$ and negative at $m=m_2$.  
This means that the form of the potential is given by Figs. \ref{potenciais}a and
\ref{potenciais}b.
 
Substituting equation (\ref{m1}) into equation (\ref{VR0}) we have
\bqn
& &V(R,m=m_1,L)=-\frac{1}{64L^4 R^6} \times \nb \\
& &\left(46 R^8 L^2+25 L^4+5 L^2 \sqrt{9 R^{16}-2 R^8 L^2+25 L^4}-32 R^6 L^4+9 R^{16}+3 R^8 \sqrt{9 R^{16}-2 R^8 L^2+25 L^4}\right), \nb \\
\eqn
and substituting equation (\ref{m2}) into equation (\ref{VR0}) we get
\bqn
& &V(R,m=m_2,L)=-\frac{1}{64L^4 R^6} \times \nb \\
& &\left(46 R^8 L^2+25 L^4-5 L^2 \sqrt{9 R^{16}-2 R^8 L^2+25 L^4}-32 R^6 L^4+9 R^{16}-3 R^8 \sqrt{9 R^{16}-2 R^8 L^2+25 L^4}\right). \nb \\
\eqn

\begin{figure}
\vspace{.2in}
\centerline{\psfig{figure=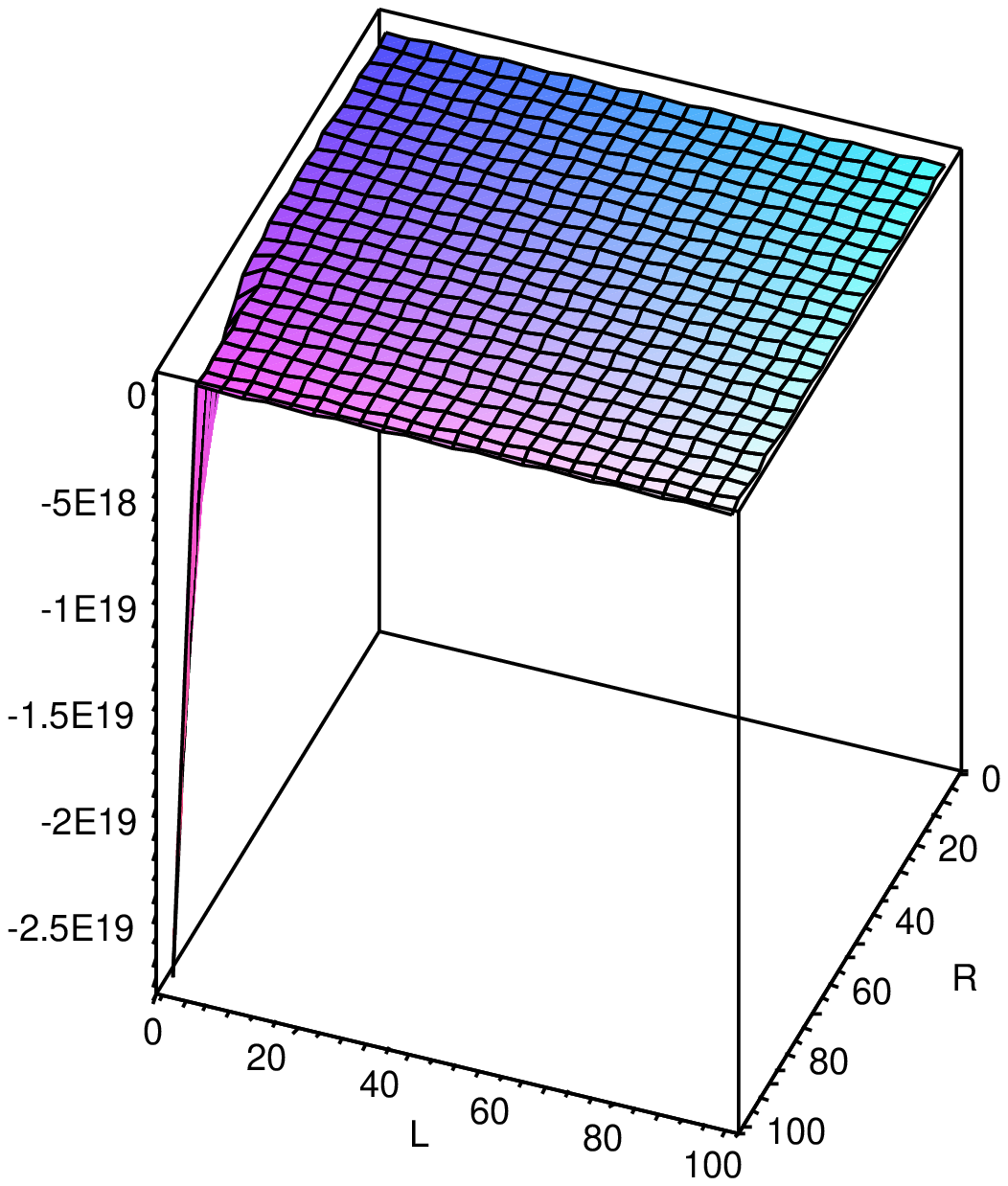,width=3.0truein,height=3.0truein}
\hskip .05in \psfig{figure=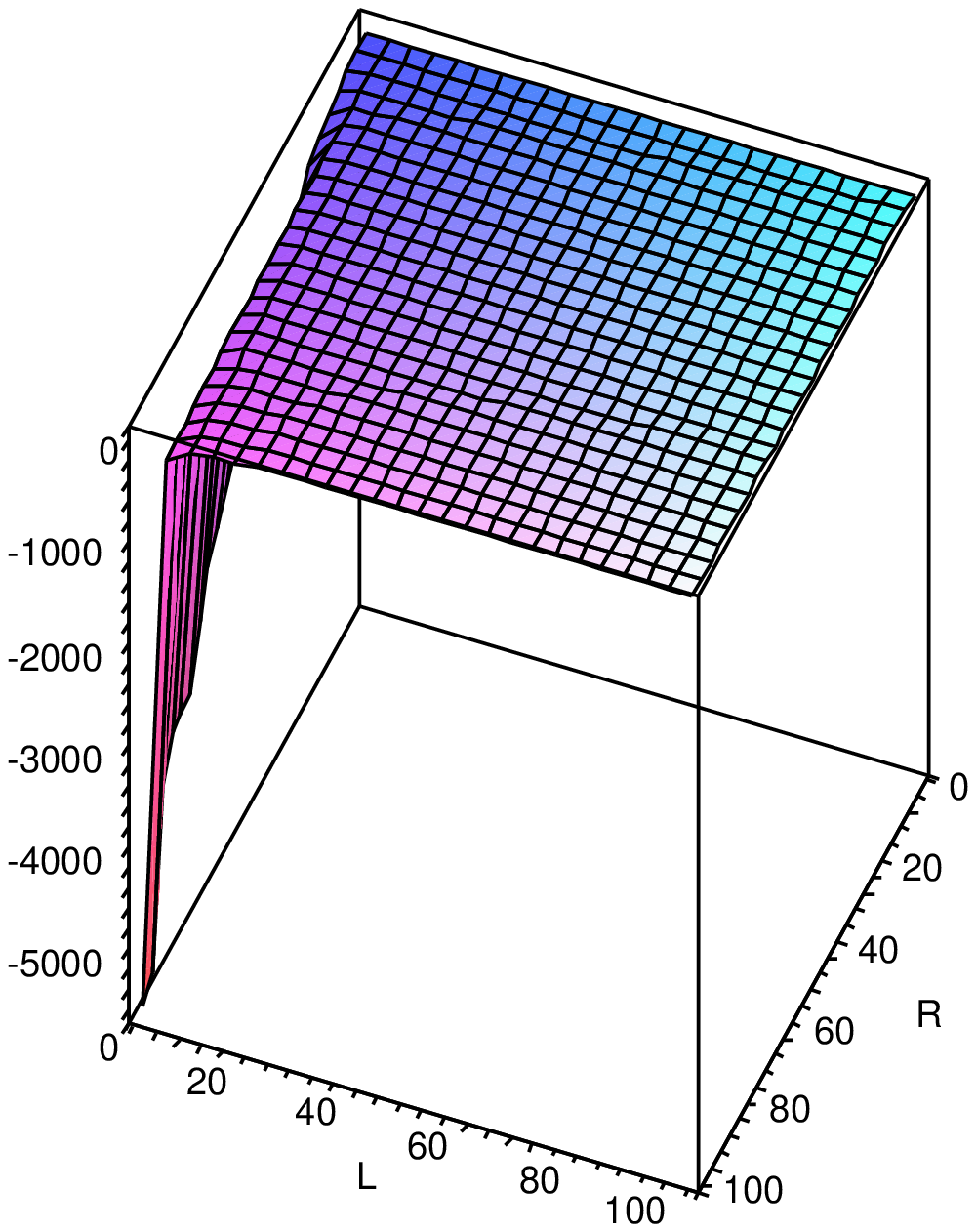,width=3.0truein,height=3.0truein}
\hskip .5in} \caption{Case $k_1 \simeq 0$. The potential $V(R,m,L)$ calculated at $m=m_1$ (left) and at $m=m_2$ (right).
We note that both potentials are negative.}
\label{fig3casek10}
\end{figure}

Thus,  from Fig. \ref{fig3casek10} we can see that the  
potential is always negative at both $m=m_1$ and  $m=m_2$.  
This means that the form of the potential is given by the lowest curves of
Figs. \ref{potenciais}a and \ref{potenciais}b, respectively.
Hence,  the collapse can either form   gravastars or  black holes.
 
\begin{figure}
\vspace{.2in}
\centerline{\psfig{figure=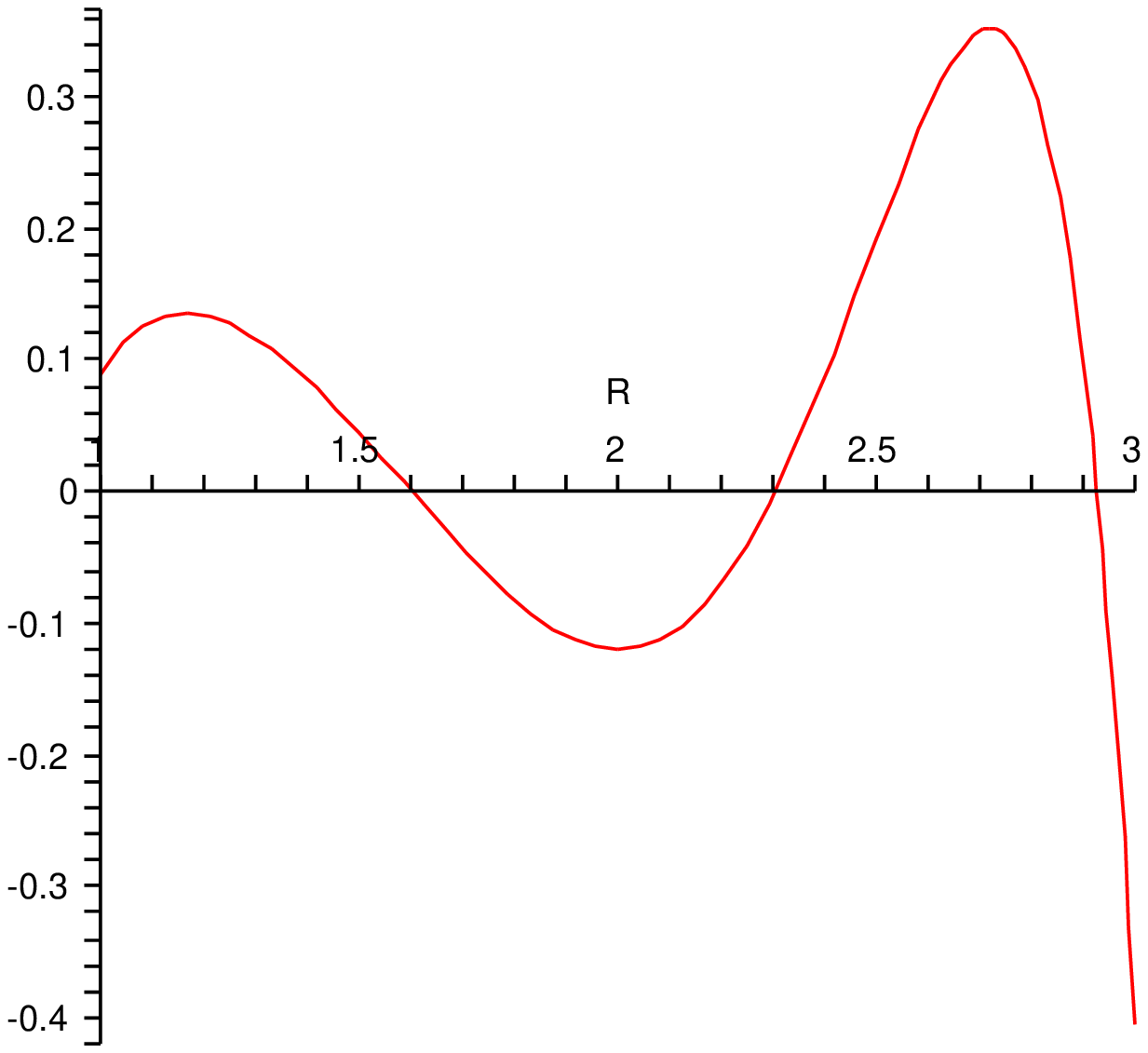,width=4.0truein,height=4.0
truein} \hskip .25in} \caption{Case $k_1 \simeq 0$. The potential 
$V(R,m,L)$ calculated at $m=m_1$, $R_c=2$ and $L_c=5$. This represents
the formation of a "bounded excursion" gravastar.}
\label{fig4casek10}
\end{figure}

\begin{figure}
\vspace{.2in}
\centerline{\psfig{figure=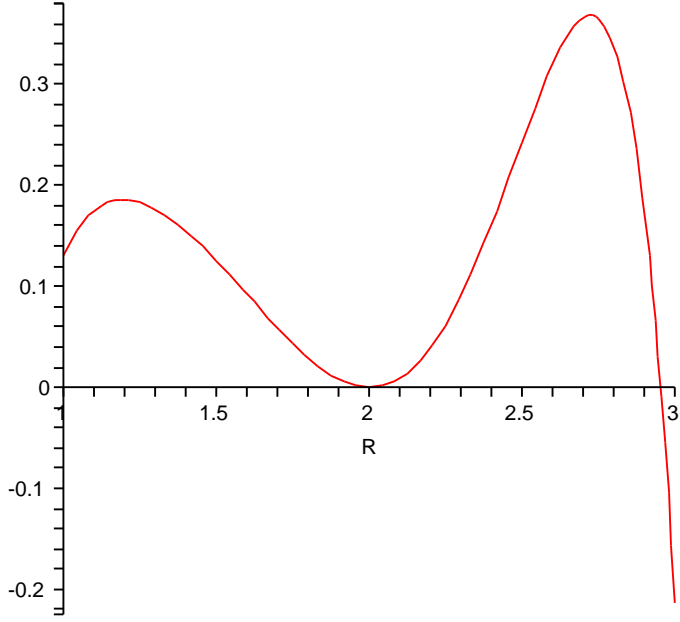,width=4.0truein,height=4.0
truein} \hskip .25in} \caption{Case $k_1 \simeq 0$. The potential 
$V(R,m,L)$ calculated at $m=m_1$, $R_c=2$ and $L_c=5.324211511$.
This represents the formation of a stable gravastar.}
\label{fig5casek10}
\end{figure}

\begin{figure}
\vspace{.2in}
\centerline{\psfig{figure=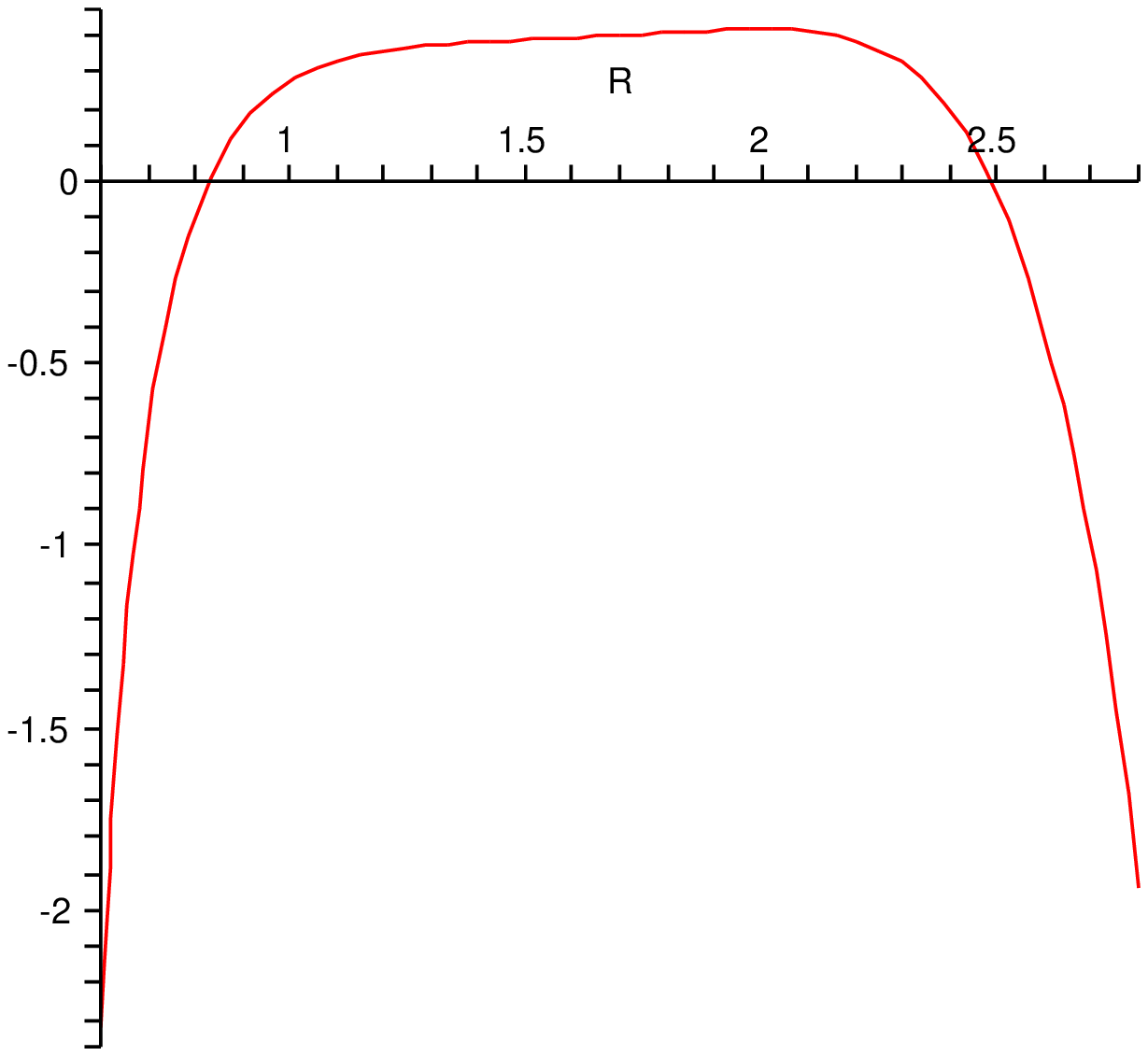,width=4.0truein,height=4.0
truein} \hskip .25in} \caption{Case $k_1 \simeq 0$. The potential 
$V(R,m,L)$ calculated at $m=m_2$, $R_c=2$ and $L_c=10$.
This represents the formation of a black hole.}
\label{fig6casek10}
\end{figure}

\section{Gravastars/Black Holes  when The Thin Shell Mass Increases}

Now, let us assume that $k_1=-2$. Then, we find that

\bq
V(R,m,L) = -\frac{1}{8L^4 R^4} \left( -4 R^4 L^4+2 L^2 R^6+4 m^2 L^4-4 m L^2 R^3+4 R^3 L^4 m+R^6+R^6 L^4 \right).
\lb{VRa}
\eq
Note again that   $\lim_{R\rightarrow 0} V(R,m,L) = \lim_{R\rightarrow \infty} V(R,m,L) =-\infty$.

The first derivative of the potential, equation (\ref{VRa}), is given by
\bq
\frac{dV}{dR}= -\frac{1}{4L^4 R^5} \left( 2 L^2 R^6+2 m L^2 R^3-2 R^3 L^4 m+R^6+R^6 L^4-8 m^2 L^4 \right).
\eq

\begin{figure}
\vspace{.2in}
\centerline{\psfig{figure=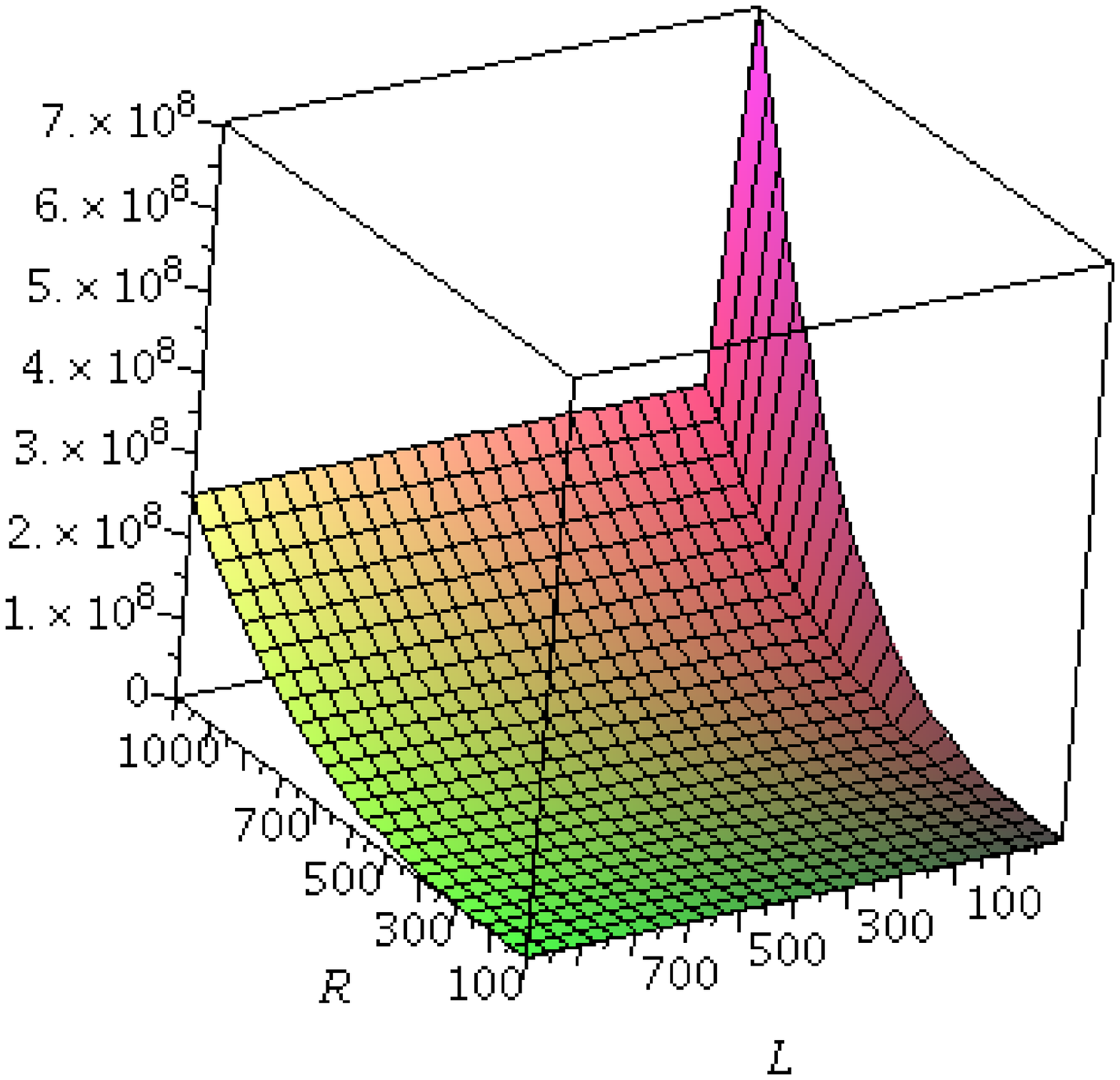,width=3.0truein,height=3.0truein}\hskip
.05in \psfig{figure=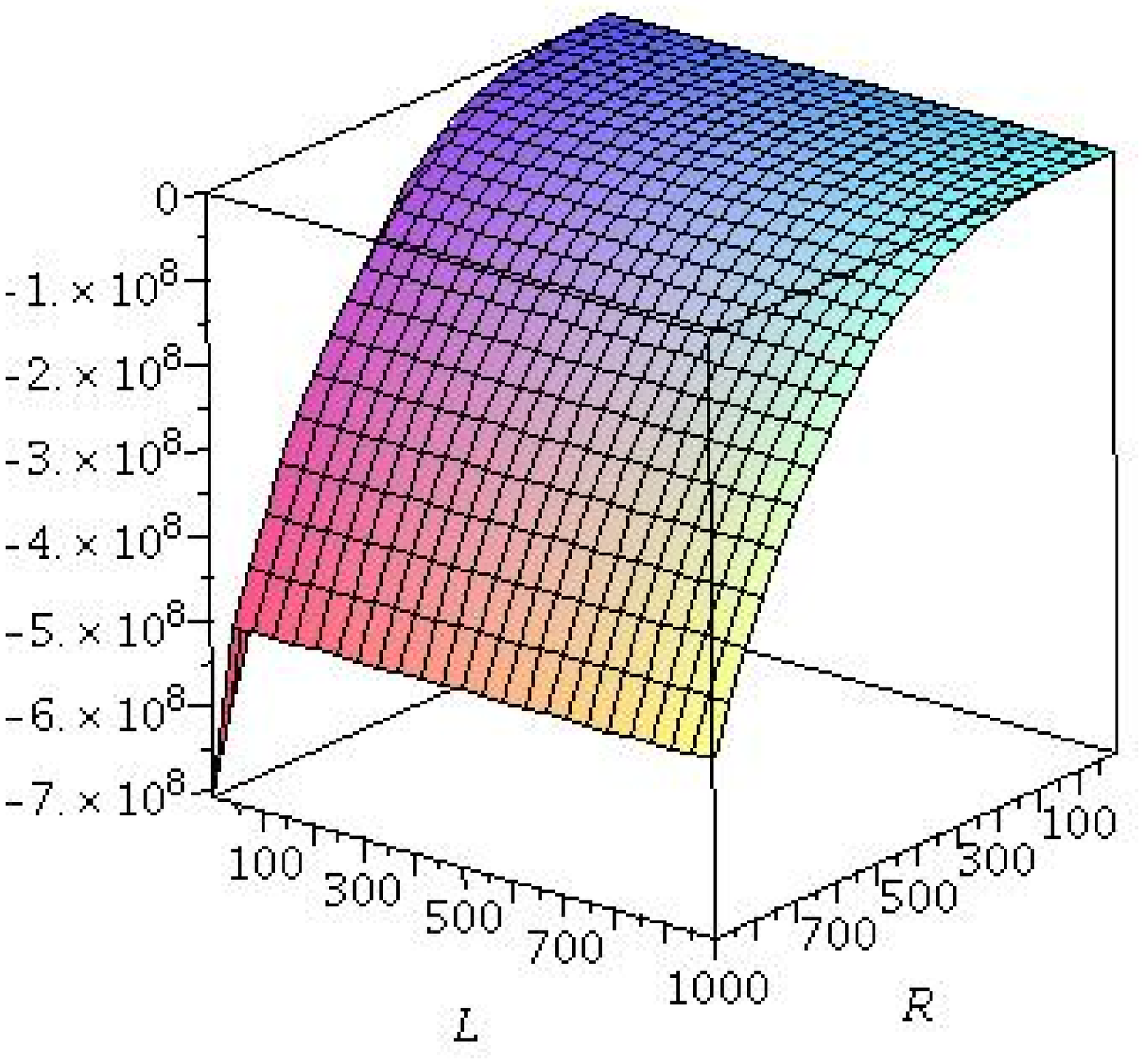,width=3.0truein,height=3.0truein}
\hskip .5in} \caption{Case $k_1=-2$. The masses $m_1$ (left) and $m_2$ (right) where the first derivative of
the potential $V(R)$ is zero.  We can note that $m_1$ is always positive and $m_2$ is always
negative.  Thus, we have only $m_1$ as solution of $\frac{dV}{dR}=0$}
\label{fig1casek1neg2}
\end{figure}

Thus,  the solutions for $\frac{dV}{dR}=0$ are
\bq
m_1 = \frac{R^3}{8L^2}\left( 1-L^2+\sqrt{9+14 L^2+9 L^4} \right)
\lb{m1a}
\eq
and
\bq
m_2 = \frac{R^3}{8L^2}\left( 1-L^2-\sqrt{9+14 L^2+9 L^4} \right).
\lb{m2a}
\eq
Note from Fig. \ref{fig1casek1neg2} that $m_2$ is always negative, while $m_1$ is always positive.  Since the mass must be always
positive, thus the unique reasonable solution for $\frac{dV}{dR}=0$
is given by $m_1$.

The second derivative of the potential is given by
\bq
\frac{d^2V}{dR^2}(R,m,L)= -\frac{1}{4R^6 L^4} \left( 2 L^2 R^6-4 m L^2 R^3+4 R^3 L^4 m+R^6+R^6 L^4+40 m^2 L^4 \right).
\lb{VR2a}
\eq

\begin{figure}
\vspace{.2in}
\centerline{\psfig{figure=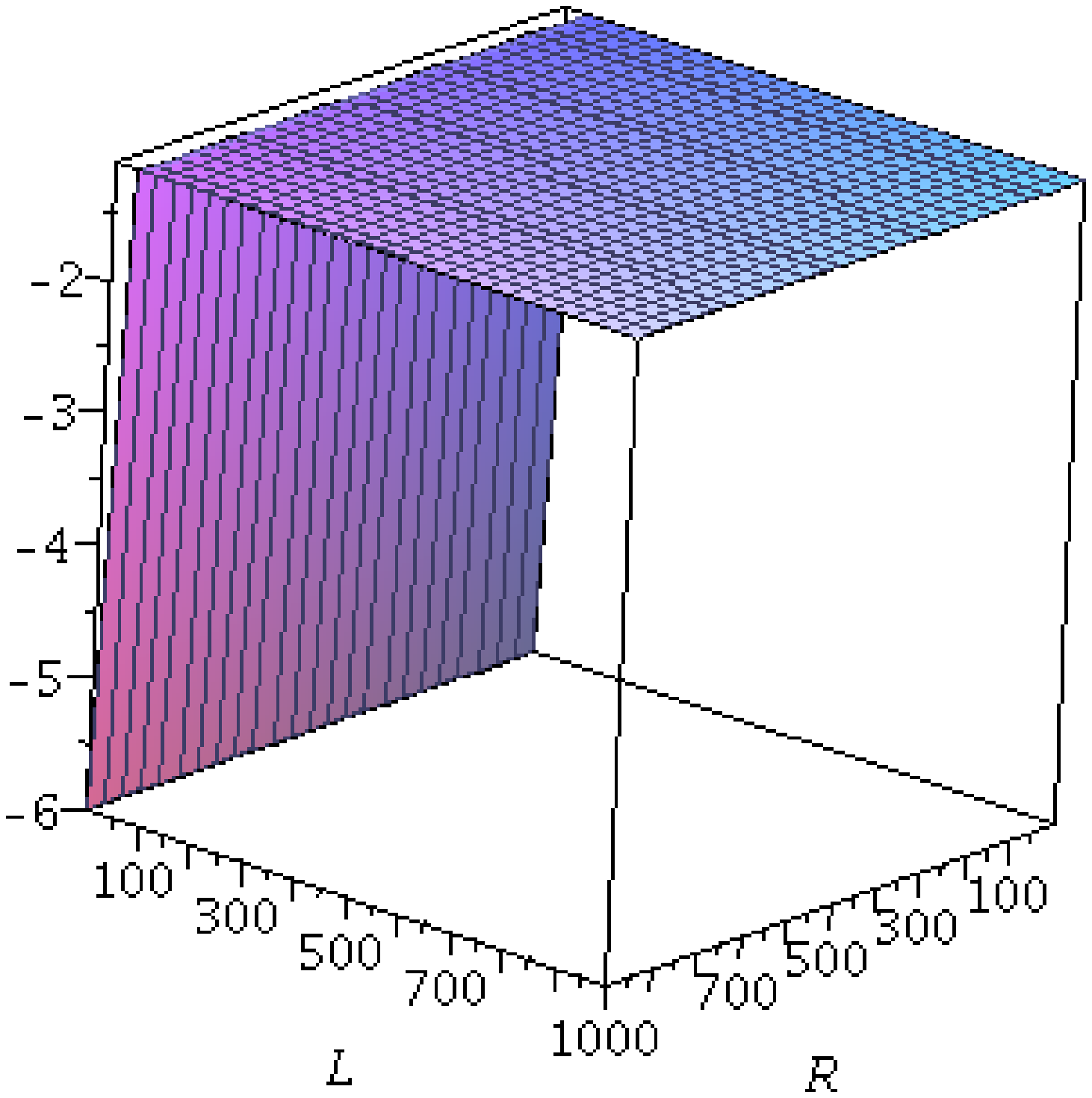,width=3.0truein,height=3.0truein}
\hskip .5in} \caption{Case $k_1=-2$. The second derivative of the potential
$\frac{d^2V}{dR^2}(R,m,L)$ calculated at $m=m_1$.
We  note that $\frac{d^2V}{dR^2}(R,m=m_1,L)$ is always negative.}
\label{fig2casek1neg2}
\end{figure}

Thus,  from Fig. \ref{fig2casek1neg2} we can see that the second derivative
of the potential is always negative at $m=m_1$.  This means that the
form of the potential is given by Fig. \ref{potenciais}b.

Substituting equation (\ref{m1a}) into equation (\ref{VR2a}) we have
\bqn
& &\frac{d^2V}{dR^2}(R,m=m_1,L)=-\frac{3}{16L^4} \times \nb \\
& &\left(14 L^2+9+\sqrt{9+14 L^2+9 L^4}+9 L^4-L^2 \sqrt{9+14 L^2+9 L^4}\right). \nb \\
\eqn

Substituting equation (\ref{m1a}) into equation (\ref{VRa}) we have
\bqn
& &V(R,m=m_1,L)=-\frac{1}{64L^4} \times \nb \\
& &\left(-32 L^4+30 R^2 L^2+9 R^2-3 R^2 \sqrt{9+14 L^2+9 L^4}+9 L^4 R^2+3 R^2 L^2 \sqrt{9+14 L^2+9 L^4}\right). \nb \\
\eqn

\begin{figure}
\vspace{.2in}
\centerline{\psfig{figure=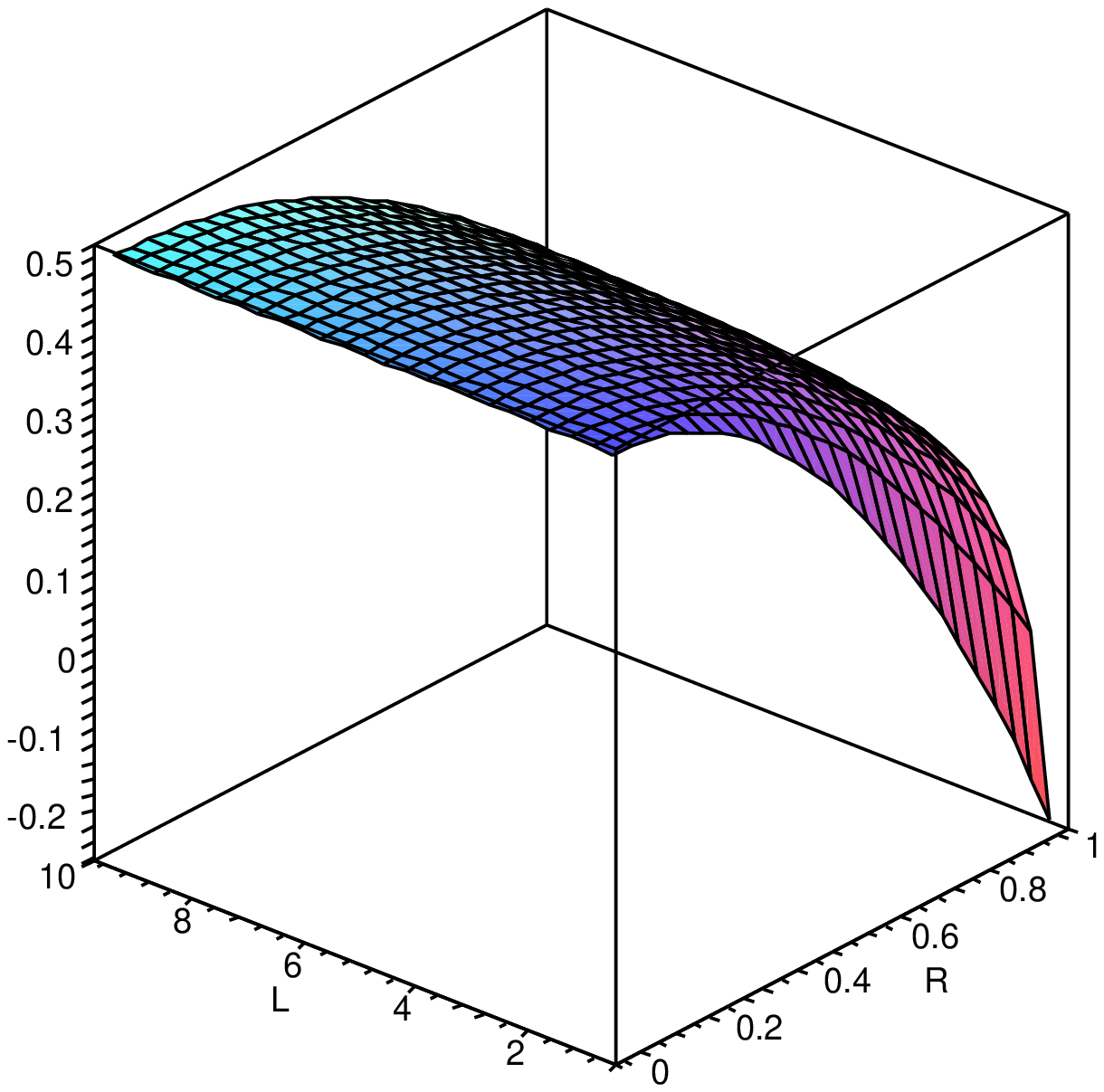,width=3.0truein,height=3.0truein}
\hskip .5in} \caption{Case $k_1=-2$. The potential $V(R,m,L)$ calculated at $m=m_1$.
We  note that $V(R,m=m_1,L)$ can be positive or negative.}
\label{fig3casek1neg2}
\end{figure}

We notice that $V(R,m=m_1,L)$ can be positive or negative, depending on the
radius $R$ and the cosmological constant $L$ (see figure \ref{fig3casek1neg2}).
This means that there exist possibilities of formation of both gravastars and  black holes.

\begin{figure}
\vspace{.2in}
\centerline{\psfig{figure=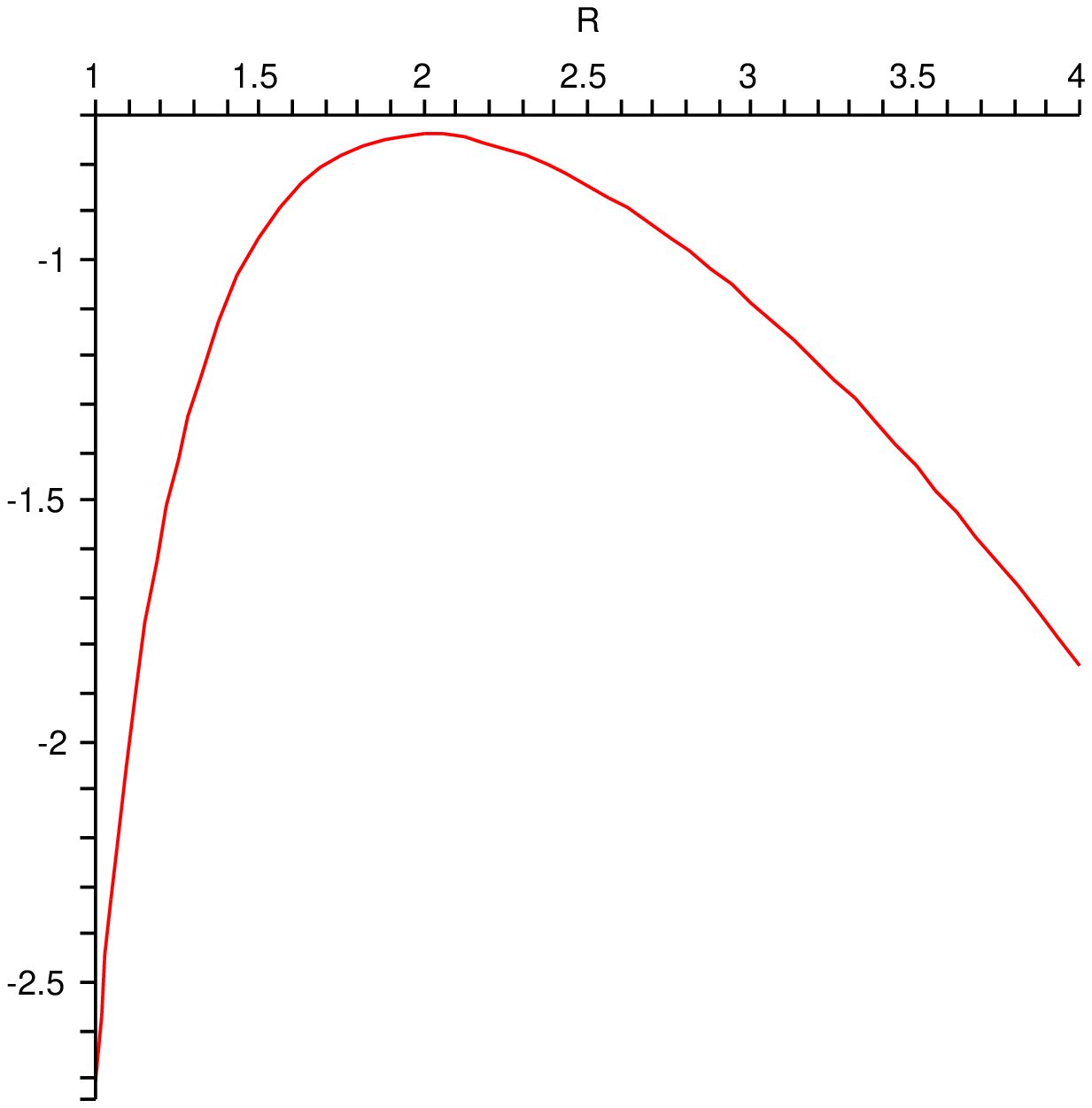,width=4.0truein,height=4.0
truein} \hskip .25in} \caption{Case $k_1 =-2$. The potential 
$V(R,m,L)$ calculated at $m=m_1$, $R_c=2$ and $L_c=10$. This
represents the formation of a black hole.}
\label{fig4casek1neg2}
\end{figure}

\subsection{Total Gravitational Mass}

In order to study the gravitational effect generated by the two components of the
gravastar, i.e., the interior de Sitter and the thin shell in the exterior region,
we need to calculate the total gravitational mass of a spherical symmetric system.
Some alternative definitions are given by \cite{Marder},\cite{Israel} and \cite{Levi}.
Here we consider the Tolman's formula for the mass, which is given by
\bq
M_{G}=\int_0^{R_0} \int_{-\pi}^{\pi} \int_0^{2\pi} \sqrt{-g}\;
T^\alpha_\alpha dr d\theta d\phi,
\eq
where $\sqrt{-g}$ is the determinant of the metric.  For the special case of a
thin shell we have
\bq
M_{G}=\int_0^{R_0} \int_{-\pi}^{\pi} \int_0^{2\pi} \sqrt{-g}\;
T^\alpha_\alpha \delta({\rr}-R_0) d{\rr} d\theta d\phi.
\eq

Thus, the Tolman's gravitational mass of the thin shell is given by
\bq
M_G^{shell}=3M,
\eq
and for the interior de Sitter (dS) spacetime we have
\bq
M_{G}^{dS}=-\frac{2}{3} \Lambda_i R_0^3.
\eq
Thus,  the de Sitter
interior presents a negative gravitational mass, since $\Lambda_i > 0$,
in agreement with its repulsive effect.

Now we can write the total Tolman's gravitational mass of the gravastar as
\bq
\label{MGtotal}
M_G^{total}=M_G^{shell}+M_G^{dS}= 3M-\frac{2}{3}\Lambda_i R_0^3.
\eq

This mass should also represent the Vaidya exterior mass ($m=M_G^{total}$)
of  the gravastar. This last equation can explain how the mass of the shell
can increase with the time.  Since $m$ must decrease with the time because
of the emission of radiation, the unique way that $M$ may increase with the
time is that the radius $R_0$   is increasing with the time.

\section{Gravastars/Black Holes when The Thin Shell Mass Decreases}

Now, let us assume that $k_1=+1$. Then, we find 

\bq
V(R,m,L) = -\frac{1}{8L^4 R^4} \left( -4 L^4 R^4+2 R^6 L^2+R^{12}-4 R^9 m L^2+4 R^6 m^2 L^4+4 L^4 m R^3+L^4 \right).
\lb{VRb}
\eq
Note again that  $\lim_{R\rightarrow 0} V(R,m,L) = \lim_{R\rightarrow \infty} V(R,m,L) =-\infty$.

The first derivative of the potential, equation (\ref{VRb}), is given by
\bq
\frac{dV}{dR}= -\frac{1}{2L^4 R^5} \left( R^6 L^2+2 R^{12}-5 R^9 m L^2+2 R^6 m^2 L^4-L^4 m R^3-L^4 \right).
\eq

Thus the solutions for $\frac{dV}{dR}=0$ are
\bq
m_1 = \frac{1}{4R^3 L^2}\left(5R^6+L^2+\sqrt{9 R^{12}+2 R^6 L^2+9 L^4}\right),
\lb{m1b}
\eq
and
\bq
m_2 = \frac{1}{4R^3 L^2}\left(5R^6+L^2-\sqrt{9 R^{12}+2 R^6 L^2+9 L^4}\right).
\lb{m2b}
\eq

\begin{figure}
\vspace{.2in}
\centerline{\psfig{figure=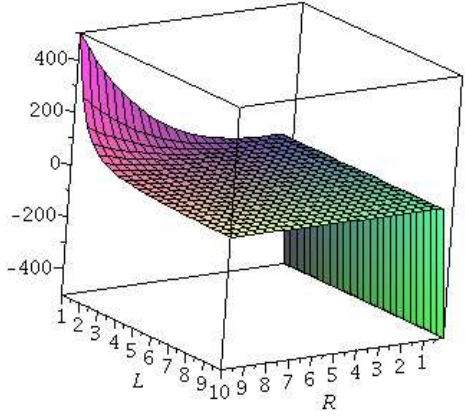,width=3.0truein,height=3.0truein}\hskip
.5in} \caption{Case $k_1=+1$. The mass $m_2$,  where the first derivative of
the potential $V(R)$ is zero.  We  note from equation (\ref{m1b}) that $m_1$ 
is always positive.  However, $m_2$ can be positive or negative.}
\label{fig1casek1pos1}
\end{figure}

We  note from equation (\ref{m1b}) that $m_1$ is always positive
and from Fig. \ref{fig1casek1pos1} that $m_2$ may be positive or negative, depending on the
radius $R$ and the cosmological constant $L$.  

The second derivative of the potential is given by
\bq
\frac{d^2V}{dR^2}(R,m,L)= -\frac{1}{2R^6 L^4} \left(R^6 L^2+14 R^{12}-20 R^9 m L^2+2 R^6 m^2 L^4+2 L^4 m R^3+5 L^4 \right).
\lb{VR2b}
\eq

\begin{figure}
\vspace{.2in}
\centerline{\psfig{figure=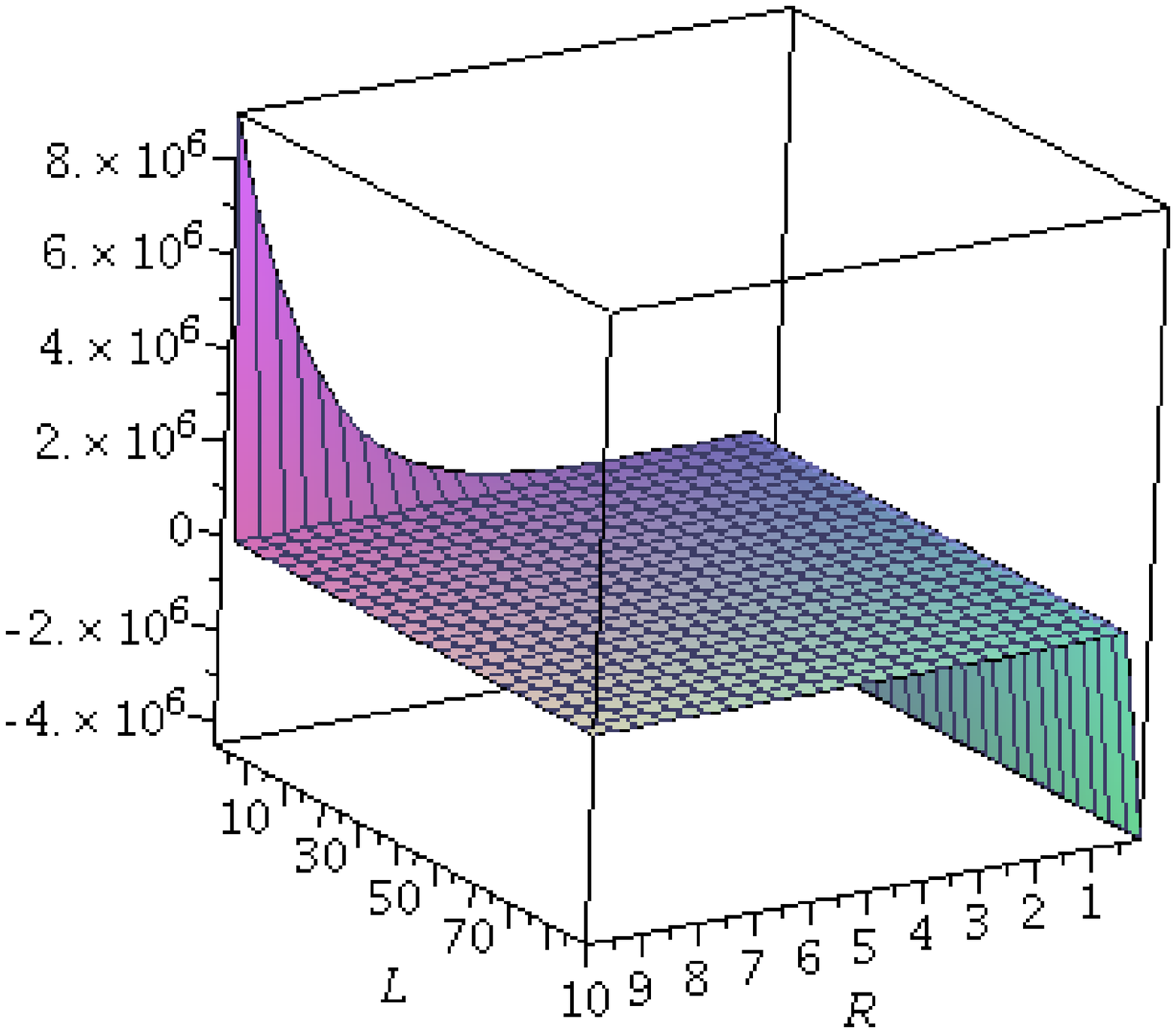,width=3.0truein,height=3.0truein}
\hskip .05in \psfig{figure=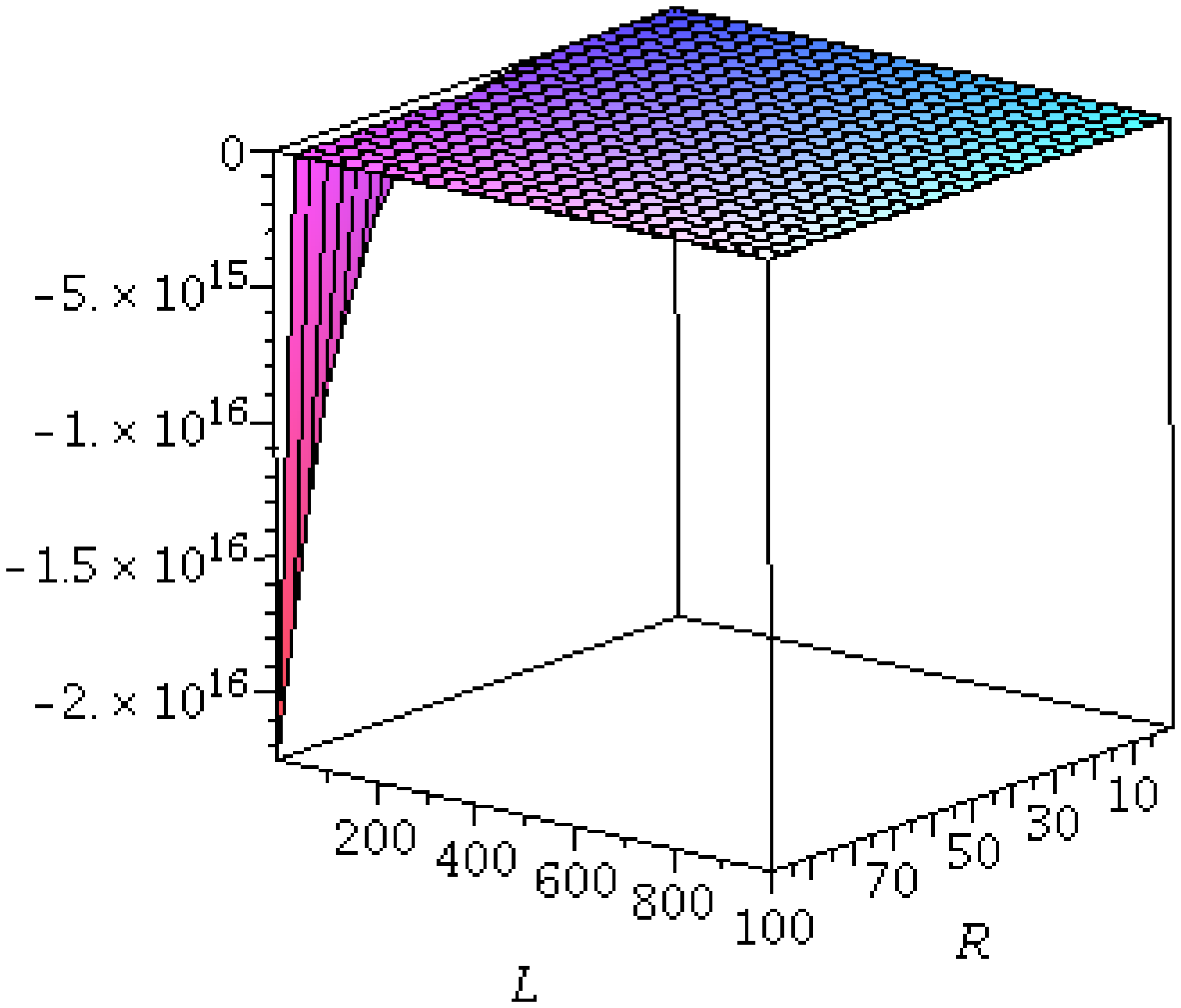,width=3.0truein,height=3.0truein}
\hskip .5in} \caption{Case $k_1=+1$. The second derivative of the potential
$\frac{d^2V}{dR^2}(R,m,L)$ calculated at $m=m_1$ (left) and at $m=m_2$ (right).
We  note that $\frac{d^2V}{dR^2}(R,m=m_1,L)$ can be positive or negative
(the frontier between the two regions is given by equation (\ref{Lf2}))
and that $\frac{d^2V}{dR^2}(R,m=m_2,L)$ is always negative.}
\label{fig2casek1pos1}
\end{figure}

Thus, we can see from Fig. \ref{fig2casek1pos1} that the second derivative
of the potential can be positive or negative at $m=m_1$, depending on the
radius $R$ and the cosmological constant $L$.  This means that the
form of the potential is given by Figs. \ref{potenciais}a and \ref{potenciais}b, respectively.
Besides, the second derivative of the potential is always negative at $m=m_2$.
This means that the form of the potential is given by Fig. \ref{potenciais}b.

Substituting equation (\ref{m1b}) into equation (\ref{VR2b}) we have
\bqn
& &\frac{d^2V}{dR^2}(R,m=m_1,L)=-\frac{3}{8L^4R^6} \times \nb \\
& &\left(9 R^{12}+5 R^6 \sqrt{9 R^{12}+2 R^6 L^2+9 L^4}-9 L^4-L^2 \sqrt{9 R^{12}+2 R^6 L^2+9 L^4}\right), \nb \\
\eqn
and substituting equation (\ref{m2b}) into equation (\ref{VR2b}) we get
\bqn
& &\frac{d^2V}{dR^2}(R,m=m_2,L)=-\frac{3}{8L^4R^6} \times \nb \\
& &\left(-9 R^{12}+5 R^6 \sqrt{9 R^{12}+2 R^6 L^2+9 L^4}+9 L^4-L^2 \sqrt{9 R^{12}+2 R^6 L^2+9 L^4}\right). \nb \\
\eqn

Solving $\frac{d^2V}{dR^2}(R,m=m_1,L)=0$, we get
\bq
L_f^2 \approx 1.816997838 R^6.
\lb{Lf2}
\eq

Substituting equation (\ref{m1b}) into equation (\ref{VRb}) we have
\bqn
& &V(R,m=m_1,L)=-\frac{1}{16L^4 R^4} \times \nb \\
& &\left(-8 L^4 R^4+18 R^6 L^2+9 R^{12}+3 R^6 \sqrt{9 R^{12}+2 R^6 L^2+9 L^4}+9 L^4+3 L^2 \sqrt{9 R^{12}+2 R^6 L^2+9 L^4}\right), \nb \\
\eqn
and substituting equation (\ref{m2b}) into equation (\ref{VRb}) we get
\bqn
& &V(R,m=m_2,L)=-\frac{1}{16L^4 R^4} \times \nb \\
& &\left(-8 L^4 R^4+18 R^6 L^2+9 R^{12}-3 R^6 \sqrt{9 R^{12}+2 R^6 L^2+9 L^4}+9 L^4-3 L^2 \sqrt{9 R^{12}+2 R^6 L^2+9 L^4}\right). \nb \\
\eqn

\begin{figure}
\vspace{.2in}
\centerline{\psfig{figure=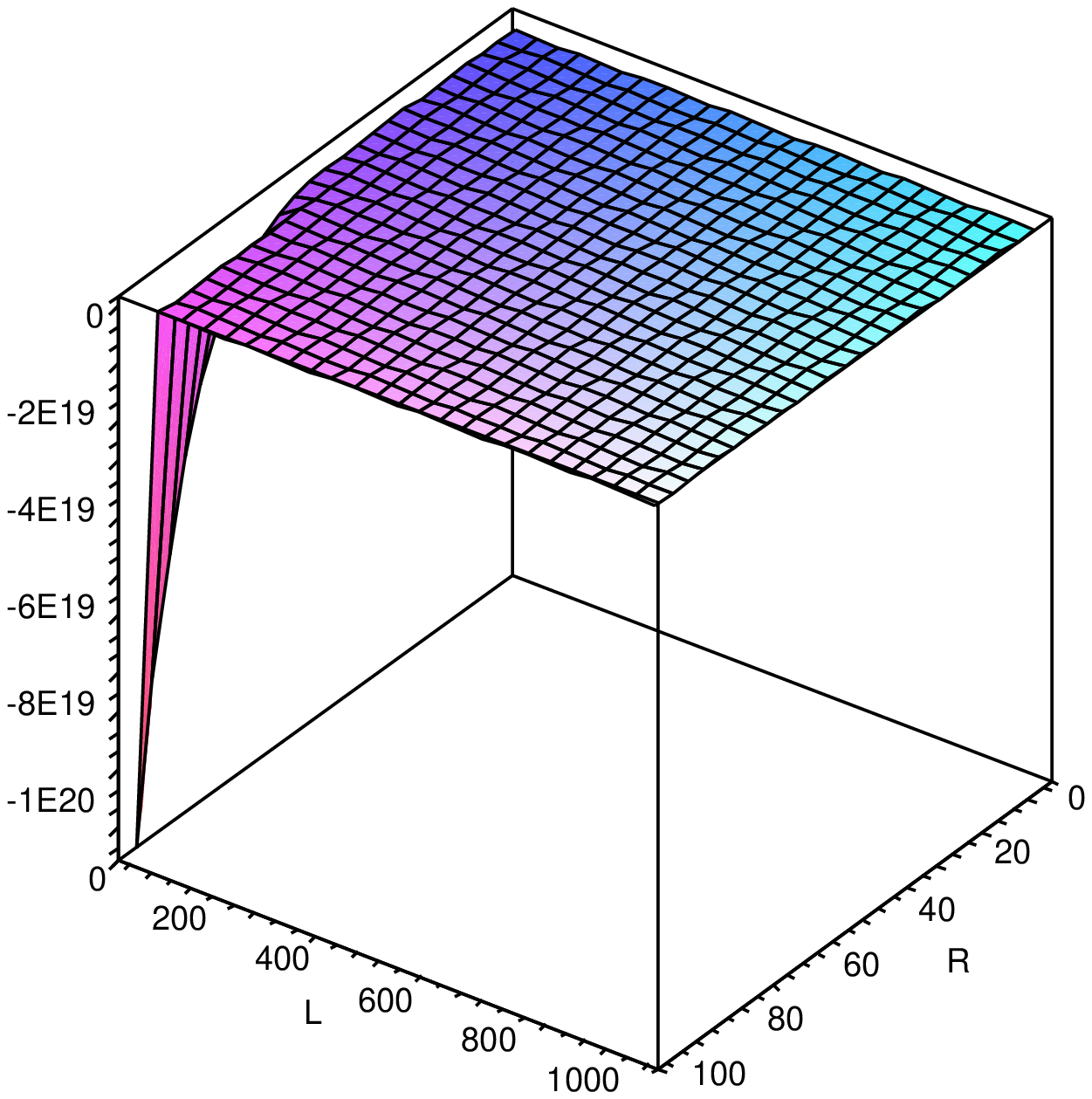,width=3.0truein,height=3.0truein}
\hskip .05in \psfig{figure=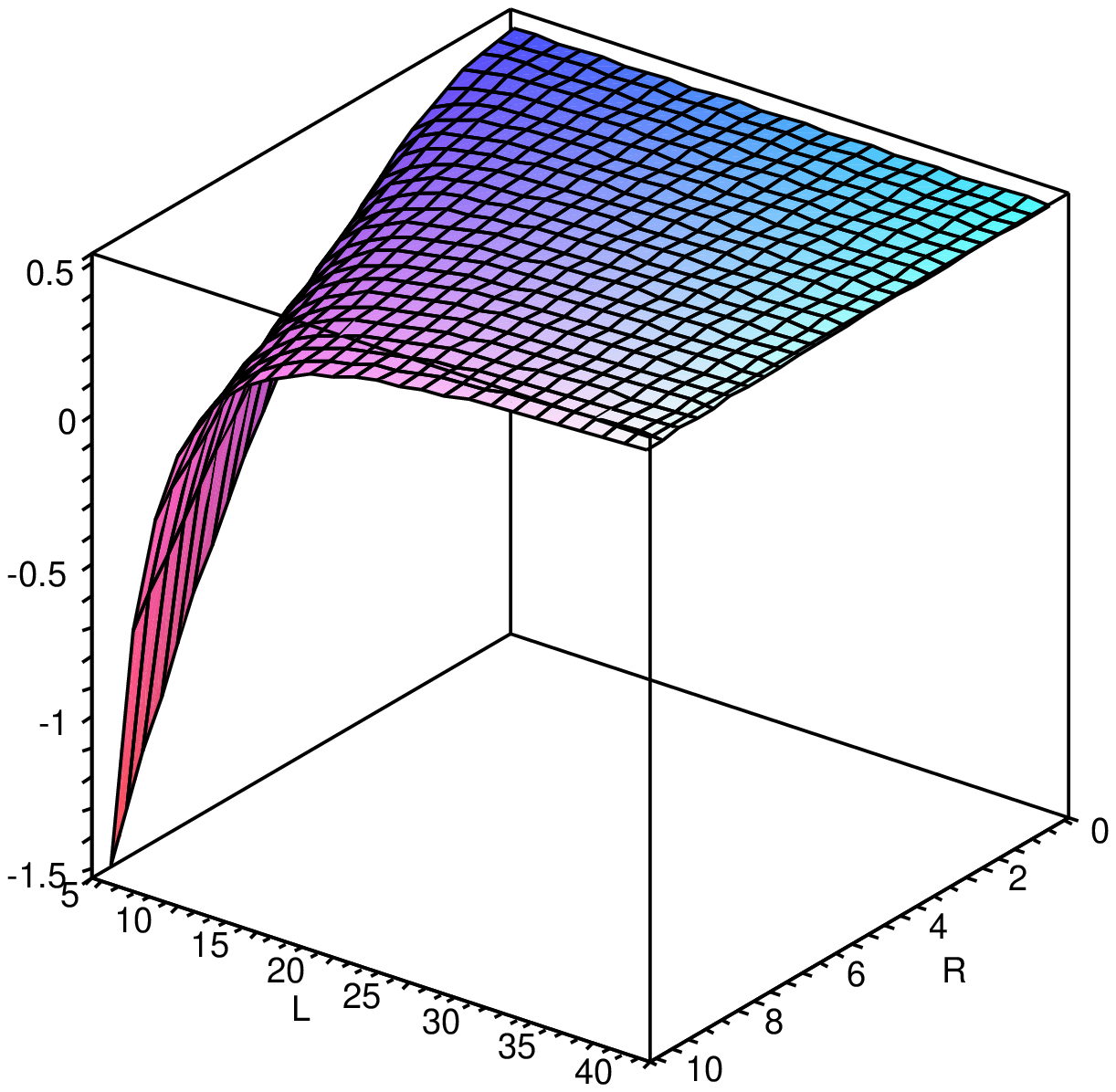,width=3.0truein,height=3.0truein}
\hskip .5in} \caption{Case $k_1=+1$. The potential $V(R,m,L)$ calculated at $m=m_1$ (left) and at $m=m_2$ (right).
We  note that $V(R,m=m_1,L)$ is always negative and that $V(R,m=m_2,L)$ can be positive or negative.}
\label{fig3casek1pos1}
\end{figure}

\begin{figure}
\vspace{.2in}
\centerline{\psfig{figure=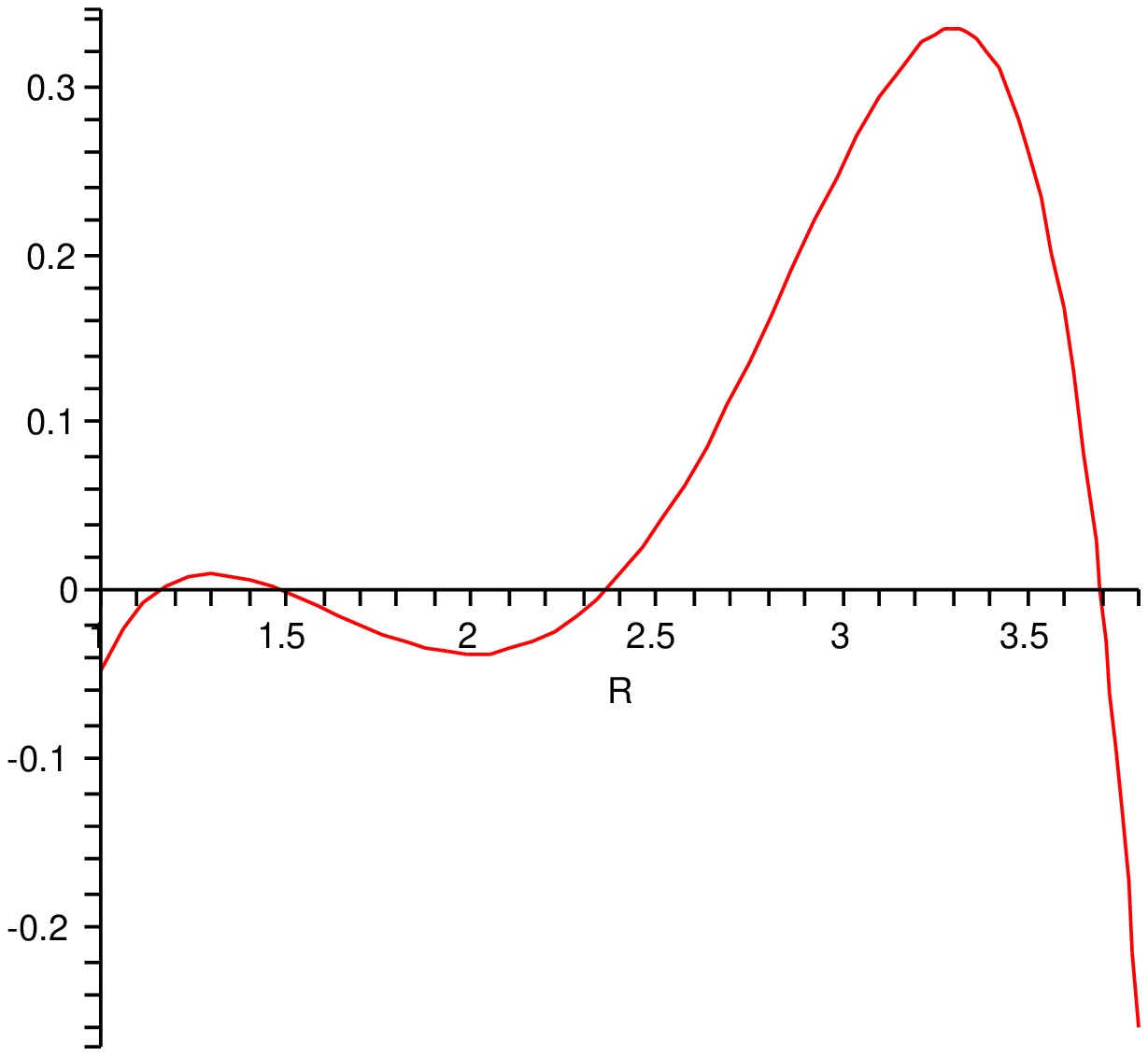,width=4.0truein,height=4.0
truein} \hskip .25in} \caption{Case $k_1 =+1$. The potential 
$V(R,m,L)$ calculated at $m=m_1$, $R_c=2$ and $L_c=5.75$.
This represents the formation of a "bounded excursion" gravastar.}
\label{fig4casek1pos1}
\end{figure}

\begin{figure}
\vspace{.2in}
\centerline{\psfig{figure=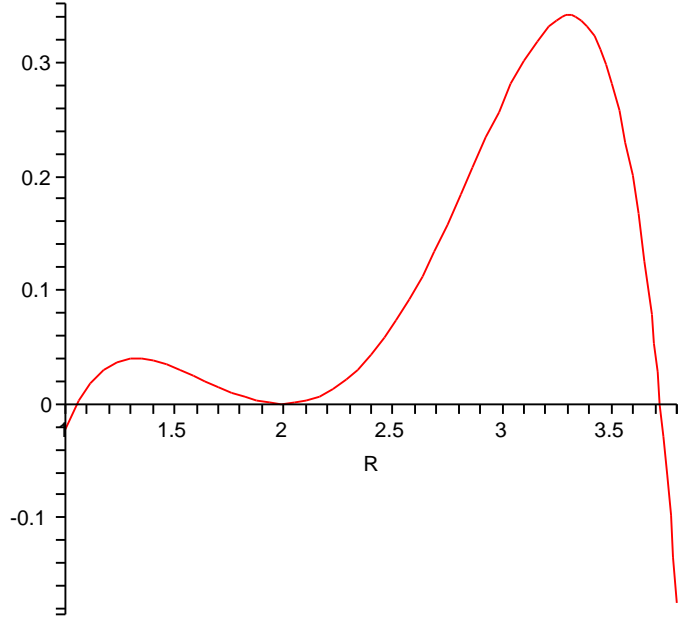,width=4.0truein,height=4.0
truein} \hskip .25in} \caption{Case $k_1 = +1$. The potential 
$V(R,m,L)$ calculated at $m=m_1$, $R_c=2$ and $L_c=5.907256196$.
This represents the formation of a stable gravastar.}
\label{fig5casek1pos1}
\end{figure}

\begin{figure}
\vspace{.2in}
\centerline{\psfig{figure=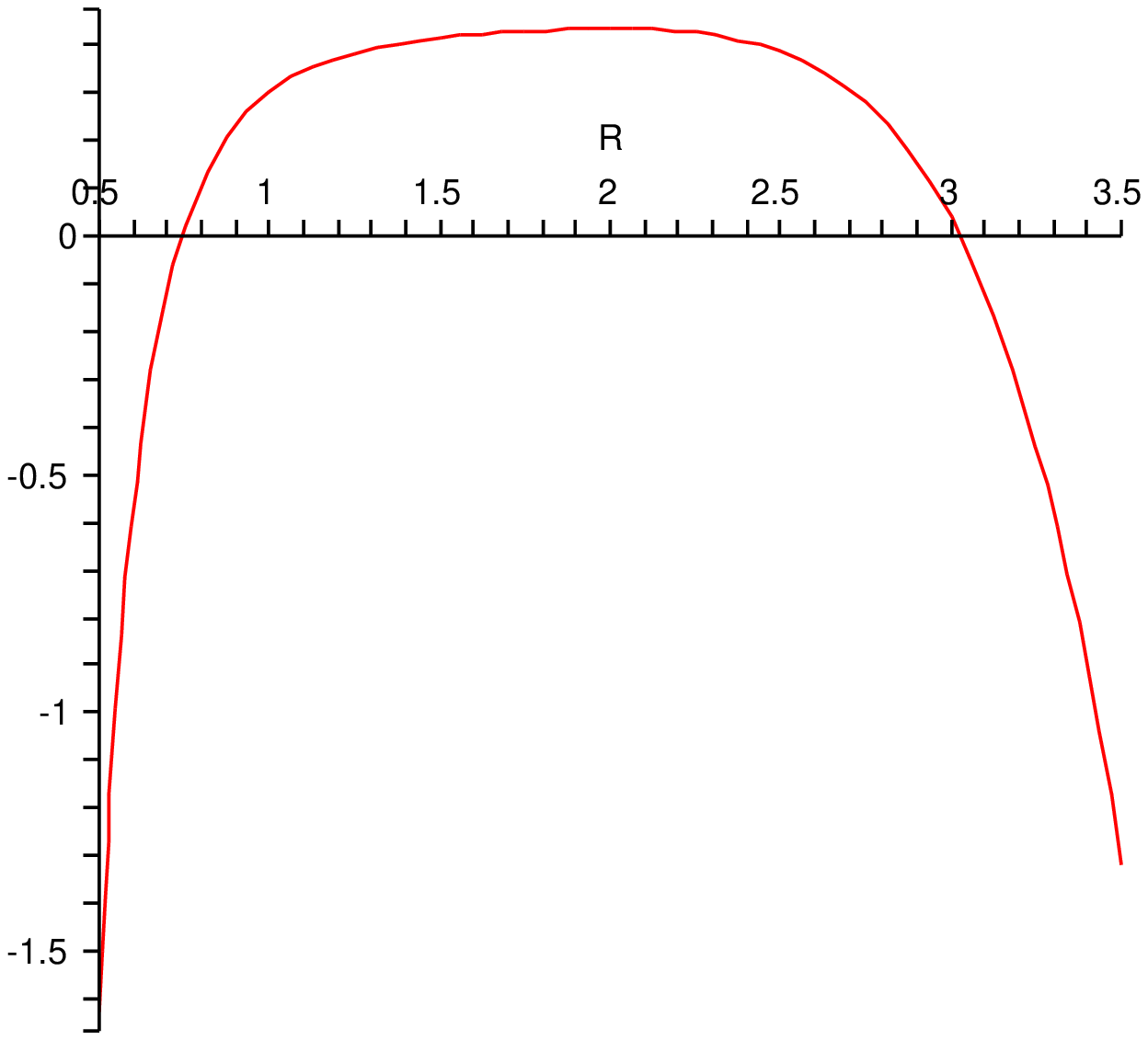,width=4.0truein,height=4.0
truein} \hskip .25in} \caption{Case $k_1 =+1$. The potential 
$V(R,m,L)$ calculated at $m=m_2$, $R_c=2$ and $L_c=5.75$.
This represents the formation of a black hole.}
\label{fig6casek1pos1}
\end{figure}

We  notice that $V(R,m=m_1,L)$ is always negative (see figure \ref{fig3casek1pos1}).
Since $V(R,m=m_2,L)$ can be positive or negative, depending on the
radius $R$ and the cosmological constant $L$,
we may have again formation of gravastars or black holes.

\section{Conclusions}

In this paper, we have studied the problem of the stability of gravastars by
constructing dynamical three-layer models  of VW \cite{VW04},
which consists of an internal de Sitter spacetime, a dynamical infinitely thin  shell of
perfect fluid with the equation of state $p = \sigma$, and an external Vaidya's spacetime.

We have shown explicitly that the final output can be a black
hole, an unstable gravastar, a stable gravastar or a "bounded excursion"
gravastar,
depending on the time evolution of the shell mass, the parameter $L$ and
the initial position $R_{0}$ of the dynamical shell. 

\begin{acknowledgments}
The financial assistance from FAPERJ/UERJ (MFAdaS) are gratefully acknowledged.
The authors (RC, MFAdaS, JFVR) acknowledges the financial support from FAPERJ (no. E-26/171.754/2000,
E-26/171.533/2002, E-26/170.951/2006, E-26/110.432/2009 and E26/111.714/2010). The authors (RC, 
MFAdaS and JFVdR) also acknowledge the financial support from Conselho Nacional de Desenvolvimento Cient\'ifico e
Tecnol\'ogico - CNPq - Brazil (no. 450572/2009-9, 301973/2009-1 and 477268/2010-2). The author (MFAdaS)
also acknowledges the financial support from Financiadora de Estudos e Projetos - FINEP - Brazil
(Ref. 2399/03). The work of AW was supported in part by DOE Grant, DE-FG02-10ER41692. 
\end{acknowledgments}

\end{document}